\documentclass[12pt]{article}
\usepackage{a4wide}
\usepackage{epsfig}
\newcommand{\eps}{\varepsilon}
\newcommand{\slk}{/\kern-6pt k}
\newcommand{\sll}{/\kern-4pt l}
\newcommand{\slp}{p\kern-5pt/}
\newcommand{\slq}{q\kern-5.5pt/}
\newcommand{\slnu}{\nu\kern-6pt/}

\newcommand{\Tr}{\mathop{\rm Tr}\nolimits}
\newcommand{\nn}{\nonumber\\}
\newcommand{\oone}{\hbox{$1\kern-2.5pt\hbox{\rm l}$}}
\newcommand{\ssigma}{\hbox{$\kern2.5pt\vrule height4pt\kern-2.5pt\sigma$}}
\newcommand{\MeV}{{\rm\,MeV}}
\newcommand{\GeV}{{\rm\,GeV}}
\newcommand{\tfrac}[2]{{\textstyle\frac{#1}{#2}}}

\newcommand{\real}{\mathop{\rm Re}\nolimits}

\newcommand{\slell}{/\kern-5pt\ell}

\begin{document}

\thispagestyle{empty} 
\begin{flushright}
MITP/15-034\\
\end{flushright}

\begin{center}
{\Large\bf Lepton-mass effects in the decays
$H \to ZZ^{\ast} \to \ell^{+} \ell^{-} \tau^{+} \tau^{-}$ and
$H \to WW^{\ast} \to \ell \nu \tau \nu_{\tau}$ }\\[1.0cm]
{\large S.~Berge$^1$, S.~Groote$^{2,3}$, J.G.~K\"orner$^3$ 
and L.~Kaldam\"ae$^2$}\\[0.3cm]
$^{1}$ Institut f\"ur Theoretische Physik, RWTH Aachen 
University, 52056 Aachen, Germany \\[.2cm]
$^2$ Loodus- ja Tehnoloogiateaduskond, F\"u\"usika Instituut,\\[.2cm]
  Tartu \"Ulikool, T\"ahe 4, 51010 Tartu, Estonia\\[7pt]
$^3$ PRISMA Cluster of Excellence, Institut f\"ur Physik,\\[.2cm]
Johannes-Gutenberg-Universit\"at,
Staudinger Weg 7, 55099 Mainz, Germany
\end{center}

\vspace{0.2cm}
\begin{abstract}\noindent
We consider $\tau$-lepton mass effects in the cascade decays 
$H\to Z(\to \ell^{+} \ell^{-})+Z^{\ast}(\to \tau^{+}\tau^{-})$ and
$H\to W^{-}(\to \ell^{-}\bar \nu_{\ell})+W^{+\ast}(\to \tau^{+}\nu_{\tau})$.
Since the scale of the problem is set by the off-shellness $q^{2}$ of the
respective gauge bosons in the limits
$(m_{\ell}+m_{\ell'})^{2} \le q^{2} \le (m_{H}-m_{W,Z})^{2}$ and not by 
$m_{W,Z}^{2}$, lepton-mass effects are non-negligible for the $\tau$ modes
in  particular close to the threshold of the off-shell decays. Lepton-mass
effects show up in the rate and in the three-fold joint angular decay 
distribution for the decays. Nonzero lepton masses lead to leptonic
helicity-flip contributions which in turn can generate novel angular
dependencies in the respective three-fold angular decay distributions.
Lepton-mass effects are more pronounced in the
$H \to Z(\to \ell\ell)Z^{\ast}(\to\tau\tau)$ mode which,
in part, is due to the fact that the ratio of lepton helicity-flip/nonflip
contributions in the decay $Z^{\ast} \to \ell^{+}\ell^{-}$ is four times
larger than in the decay $W^{+\ast}\to \ell^{+}\nu$. Overall the inclusion of
$\tau$ mass effects leads to a $3.97 \%$ reduction in the leptonic
$H \to ZZ^{\ast}$ rate. Lepton mass effects are quite pronounced for $q^{2}$
values from threshold up to $\sim 200\GeV^{2}$.  For example, at
$q^{2}=50\GeV^{2}$ the transverse--longitudinal--scalar helicity composition
of the off-shell $Z$--boson changes from $0.06:0.94:0$ to $0.04:0.65:0.31$ for
the $\tau$ lepton. This has observational consequences for the angular decay
distributions of the final-state leptons. We also briefly consider the
corresponding off-shell -- off-shell decays
$H\to Z^{\ast}(\to \ell^{+}\ell^{-})+Z^{\ast}(\to \tau^{+}\tau^{-})$ and
$H\to W^{-\ast}(\to \ell^{-}\bar \nu_{\ell})
+W^{+\ast}(\to \tau^{+}\nu_{\tau})$. 
\end{abstract}

\newpage

\section{Introduction}
We consider lepton-mass effects in the off-shell decays of gauge bosons
in the processes $Z^{\ast}\to \tau^{+}\tau^{-}$ and  
$W^{+\ast} \to \tau^{+}\nu_{\tau}$ where the off-shell gauge bosons
$W^{+\ast},Z^{\ast}$ are produced in the Higgs decays $H \to ZZ^{\ast}$,
$W^{-}W^{+\ast}$. In the $H \to ZZ^{\ast}$ case the corresponding 
$\ell=e,\mu$ modes have recently been observed at the LHC and are therefore 
adequately dubbed ``Higgs discovery channels''~\cite{:2012gk,:2012gu}.
Further evidence on these decays has been presented in
Ref.~\cite{Chatrchyan:2012jja}. The quantum numbers of the Higgs boson have
been pinned down by an angular analysis of the four leptons in the final state
to be $J^{P}=0^{+}$ both in the leptonic $H \to ZZ^{\ast}$ 
mode~\cite{Chatrchyan:2012jja,Chatrchyan:2013mxa,Aad:2013xqa} as well as in
the leptonic $H \to W^{-}W^{+\ast}$ mode~\cite{Aad:2015rwa}. On the theoretical
side there have been a number of papers analyzing the quantum numbers of the
Higgs boson through an angular analysis of the four-lepton final state among
which are Refs.~\cite{Choi:2002jk,Kovalchuk:2008zz,Gao:2010qx,DeRujula:2010ys,%
Bolognesi:2012mm,Avery:2012um,Sun:2013yra,Buchalla:2013mpa,Beneke:2014sba,%
Gainer:2014hha, Modak:2013sb,Bhattacherjee:2015xra,Zagoskin:2015sca}. The
physics of the Higgs boson in all its aspects has been nicely reviewed in
three recent papers~\cite{Ellis:2015daa,Ellis:2015tba,Djouadi:2015haa}.

Off-shell effects in the decays involving massive leptons will lead to 
additional scalar and scalar--longitudinal interference contributions
well familiar from neutron beta decay, the semileptonic decay
$\Xi^{0}\to\Sigma^{+}\mu^{-}\bar{\nu}_{\mu}$~\cite{Kadeer:2005aq}, or from the
decays $B \to D^{(\ast)}\tau \nu_{\tau}$~\cite{Korner:1989ve,Korner:1989qb}
and $\Lambda_{b} \to \Lambda_{c} \tau \nu_{\tau} $~\cite{Gutsche:2015mxa}.
The scalar and scalar--longitudinal interference contributions are quadratic 
in the lepton masses and can thus be neglected at the scale $m_{W,Z}^{2}$. 
However, for the off-shell decays $H \to ZZ^{\ast}, W^{-}W^{+\ast}$ 
the scale is not 
set by $m_{W,Z}^{2}$ but by the off-shellness of the
respective gauge bosons which extends from threshold 
$q^{2}=(m_{\ell}+m_{\ell'})^{2}$ 
(maximal recoil point) to the zero recoil point at 
$q^{2} = (m_{H}-m_{W,Z})^{2}$, i.e. one has 
\begin{equation}
(m_{\ell}+m_{\ell'})^{2} \le q^{2} \le (m_{H}-m_{W,Z})^{2}\,.
\end{equation}
One will therefore have to carefully consider $\tau$-lepton mass effects   
particularly in the $q^{2}$ region close to threshold given by
$q^{2}= 4m_{\tau}^{2}$ and $q^{2}=m_{\tau}^{2}$ for the leptonic modes in the
decays $H \to ZZ^{\ast}$ and $H \to WW^{\ast}$, respectively. Lepton-mass
effects reduce the overall rate relative to the zero lepton-mass case. In
addition, lepton-mass effects lead to leptonic helicity-flip contributions
which in turn can generate novel angular dependencies in the respective
angular decay distributions. These angular dependencies can mimic new angular
terms introduced by higher dimension effective coupling
terms~\cite{Buchalla:2013mpa,Beneke:2014sba,Gainer:2014hha} or non-SM 
$(HVV)$ coupling
terms~\cite{Modak:2013sb,Bhattacherjee:2015xra,Zagoskin:2015sca}. 
$\tau$-lepton mass effects should therefore not be neglected if one is aiming
for high precision physics in the Higgs sector.\footnote{Lepton-mass effects
in the rate $H \to W\ell\nu$ and $H \to Z \ell\ell$ are also taken into 
account in Ref.~\cite{Kniehl:2012rz}.}

Our paper is structured as follows. After this introductory section, in Sec.~2
we present a general formula for the three-fold angular angular decay
distribution for the on-shell -- off-shell decays
$H \to VV^{\ast} \to \ell\ell\ell\ell$. The angular decay distribution is
obtained using helicity methods. In Sec.~3 we discuss lepton-mass effects in
the decay $H \to ZZ^{\ast} \to \ell\ell\tau\tau$ and their effect on the rates
and the angular decay distributions. We do the same in Sec.~4 for the decays
$H \to W^{-}W^{+\ast} \to \ell\nu_\ell\tau\nu_\tau$. In Sec.~5 we summarize
our results and conclude with some general remarks. Some technical material
regarding helicity amplitudes is relegated to the Appendices. In Appendix~A we
list the helicity amplitudes for the $H \to VV^{\ast}$ transitions. The
helicity representation of the lepton tensors in the neutral- and
charged-current cases can be found in Appendix~B and~C, respectively.

\section{General formalism}
The three-fold angular decay distribution in the cascade decays
$H \to VV^{\ast} \to \ell\ell\ell\ell,\quad V=Z,W$ can be derived from the
covariant contraction of the on-shell and off-shell lepton tensors
$L^{(p)}_{\mu\nu}$ and $L^{(q)}_{\mu\nu}$ with the $(HVV)$ Higgs coupling
$H_{\alpha\beta}$ where the vertices are connected by the propagator
projectors $P_{1}^{\alpha\mu}$ (spin 1) and $P_{0\oplus1}^{\nu\beta}$
(spin $0\oplus1$). One has
\begin{equation}\label{angdis1}
W(\theta_{p},\theta_{q},\chi)=
H_{\alpha\alpha'}\,P_{1}^{\alpha\mu}(p)\,P_{0\oplus1}^{\alpha'\mu'}(q)\,
L^{(p)}_{\mu\nu}(p)\,L^{(q)}_{\mu'\nu'}(q)\,P_{1}^{\nu\beta}(p)
\,P_{0\oplus1}^{\nu'\beta'}(q)\,H_{\beta\beta'}^{\ast}
\end{equation}
where, in the Standard Model (SM), $H_{\alpha\alpha'}=g_{\alpha\alpha'}$. We
denote the on-shell and off-shell momenta of the gauge bosons by $p$
and $q$. In the unitary gauge the on-shell spin-1 propagator
$P_{1}^{\alpha\mu}$ ($p^{2}=m_{V}^{2}$) and the off-shell propagator
$P_{0\oplus1}^{\nu\beta}$ ($q^{2}\neq m_{V}^{2}$) read
\begin{equation}\label{offshell}
P_{1}^{\alpha\mu}(p)=-g^{\alpha\mu}+\frac{p^{\alpha}p^{\mu}}{p^{2}},\qquad
P_{0\oplus1}^{\nu\beta}(q)=-g^{\nu\beta}+\frac{q^{\nu}q^{\beta}}{m_{V}^{2}}.
\end{equation}
Note that in the unitary gauge\footnote{The choice of the unitary gauge is
mandatory to obtain a gauge-independent result. This can be seen by
considering a general covariant $R_\xi$ gauge where one has to consider
Goldstone boson exchange in addition to gauge boson exchange. In the coupling
to the final state fermion pair the gauge parameter $\xi$ cancels between
the Goldstone and gauge boson contributions, resulting in the unitary gauge
propagator. This has been explicitly demonstrated for fermion--fermion
scattering~\cite{Peskin:1995ev} and for the decay
$t\to b +W^{+\ast}$~\cite{Korner:2014bca}.} the off-shell propagator
$P_{0\oplus1}^{\nu\beta}(q)$ contains a spin-1 and a spin-0 piece. This can be
seen by splitting the off-shell gauge propagator in Eq.~(\ref{offshell}) into
its spin-1 and spin-0 components according to
\begin{equation}\label{split}
P_{0\oplus1}^{\nu\beta}(q)=-g^{\nu\beta}+\frac{q^{\nu}q^{\beta}}{m_{V}^{2}}
=\bigg(\underbrace{-g^{\nu\beta}+\frac{q^{\nu}q^{\beta}}{q^{2}}}_{\rm spin\,1}
\bigg)-\underbrace{\frac{q^{\nu}q^{\beta}}{q^{2}}F_{S}(q^{2})}_{\rm spin\,0},
\end{equation} 
where
\begin{equation}\label{scalar}
F_{S}(q^{2})=\left(1-\frac{q^{2}}{m_{V}^{2}}\right).
\end{equation}
In the zero lepton-mass approximation one has $q^{\mu}L_{\mu\nu}=0$ and
therefore the spin-0 piece in Eq.~(\ref{split}) does not contribute and can be
dropped when evaluating Eq.~(\ref{angdis1}). This is always a good
approximation for $\ell=e,\mu$ but no longer a good approximation for
$\ell=\tau$. An interesting observation concerns the spin-0 contribution. 
Taken together with the propagator pole proportional to
$(q^2-m_V^2)^{-1}$, the contribution of the spin-0 piece can be seen to be
proportional to a contact interaction of the form $(HV\psi\bar\psi)$ with a
$q^2$-dependent coupling when one sets $\Gamma_Z=0$.

\begin{figure}
\begin{center}
\epsfig{figure=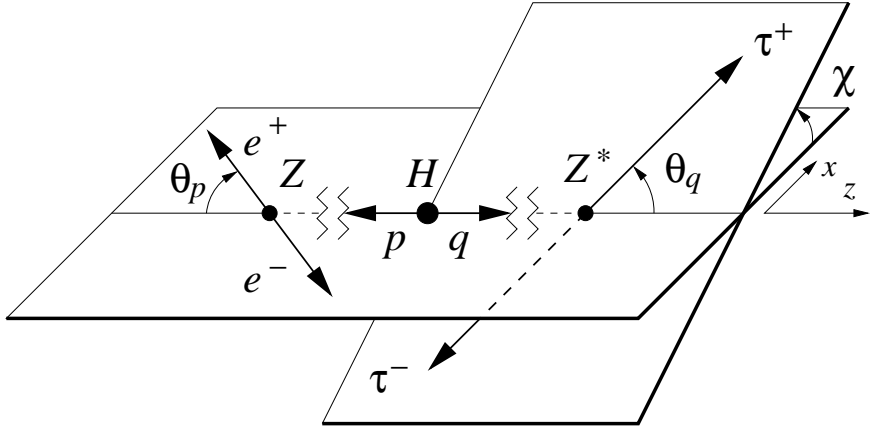, scale=0.7}
\end{center}
\caption{\label{planeszz}Definition of the momenta $p$ and $q$, the polar
angles $\theta_{p}$ and $\theta_{q}$, and the azimuthal angle $\chi$ in the
cascade decay $H\to Z(\to e^{+}e^{-})+Z^{\ast}(\to \tau^{+}\tau^{-})$}
\end{figure}

Technically there are two routes to obtain angular decay distributions
from Eq.~(\ref{angdis1}). In the first route one parametrizes the four-vectors
of the problem in terms of the five phase-space variables $p^{2}=m_{V}^{2}$,
$q^{2}$, $\cos\theta_{p}$, $\cos\theta_{q}$ and $\chi$ (cf.\
Figs.~\ref{planeszz} and~\ref{planesww}). The covariant evaluation of the
Lorentz-invariant expression~(\ref{angdis1}) leads to a number of scalar
products of momenta that are defined in different reference frames. When doing
the requisite contractions, the four-momenta have to be boosted to a common
reference frame as e.g.\ described in Ref.~\cite{Buchalla:2013mpa} for the
decay $H \to Z\ell\ell$ and in Ref.~\cite{Cappiello:2011qc,Gevorkyan:2014waa}
for the decay $K^{\pm}\to \pi^{\pm}\pi^0e^+e^-$. One then arrives at the
desired three-fold joint angular decay distribution.

A second, perhaps more intelligent route, is to use an analysis in terms of
helicity amplitudes. The advantage of the helicity method is that the origin
of the angular factors multiplying the helicity structure functions can be
straightforwardly identified. The angular factors can be seen to arise from
the transformation properties of the helicity amplitudes under the action of
the rotation group.  

In order to transform to the helicity representation of the covariant form in
Eq.~(\ref{angdis1}) one makes use of the completeness relation for the spin-1
on-shell and off-shell polarization vectors. The on-shell and off-shell
propagator can be expanded according to~\cite{Korner:1989qb,Korner:1987kd}
\begin{equation}
\label{complete1}
P_{1}^{\alpha\mu}(p)=-g^{\alpha\mu}+\frac{p^{\alpha}p^{\mu}}{p^2}
=\sum_{\lambda_{V}=\pm1,0}\bar \eps^{\alpha}(\lambda_{V})\bar 
\eps^{\ast\,\mu}(\lambda_{V})
\end{equation} 
($p^{2}=m_{V}^{2}$) and
\begin{equation}
\label{complete2}
P_{0\oplus1}^{\mu'\alpha'}(q)=-g^{\mu'\alpha'}
+\frac{q^{\mu'}q^{\alpha'}}{m_{V}^{2}}=
-\sum_{\lambda_{V^{\ast}}=t,\pm1,0}\eps^{\mu'}(\lambda_{V^{\ast}})
\eps^{\ast\,\alpha'}(\lambda_{V^{\ast}})
\,\hat g_{\lambda_{V^{\ast}}\lambda_{V^{\ast}}}.
\end{equation} 
Note that there is an additional spin-0 degree of freedom propagating in the
off-shell propagator in Eq.~(\ref{complete2}). We shall specify this spin-0
degree of freedom by assigning the label $\lambda_{V}=t$ ($t$ for
time-component) to this mode. According to the separation in Eq.~(\ref{split})
the ``$t$'' mode carries the weight $F_{S}=(1-q^{2}/m_{V}^{2})$ which finally
leads to $\hat g_{\lambda_{V^{\ast}}\lambda_{V^{\ast}}}
=diag\,\{F_{S},-1,-1,-1\}$ in Eq.~(\ref{complete2}). The four polarization
four-vectors $\eps^{\mu}(t,\pm1,0)$ will be specified in Appendix~A.

In low-energy calculations such as neutron $\beta$ decay or in the
semileptonic bottom-hadron decays one usually drops the term proportional to
$(q^{2}/m^{2}_{V})$ in Eq.~(\ref{scalar}) since one has $q^{2}\ll m_{V}^{2}$.
However, in the present application the factor $(q^{2}/m_{V}^{2})$ can become
as large as $30\%$ at the zero recoil point and can therefore not be
neglected.\footnote{In muon decays $\mu^{-}\to e^{-}+\mu_{\mu}+\bar\nu_{e}$
where one is aiming for ultrahigh precision, the importance of the
$q^{2}/m_{W}^{2}$ contributions have been discussed in the
literature~\cite{Fischer:2002hn,Fael:2013pja}.}

With the help of the completeness relations~(\ref{complete1})
and~(\ref{complete2}) the covariant form of the angular decay
distribution~(\ref{angdis1}) can be cast into a representation in terms of
helicity components. One has
\begin{equation}\label{angdis2}
W(\theta_{p},\theta_{q},\chi)=\sum_{\lambda_{V},\lambda'_{V}=
\pm1,0 \atop \lambda_{V^{\ast}},\lambda'_{V^{\ast}}=t,\pm1,0}
(-F_{S})^{2-J-J'}L_{\lambda_{V}\,\lambda'_{V}}^{(p)}(\cos\theta_{p})
H_{\lambda_{V},\lambda_{V^{\ast}}}H^{\ast}_{\lambda'_{V},\lambda'_{V^{\ast}}}
L^{(q)}_{\lambda_{V^{\ast}}\,\lambda'_{V^{\ast}}}(\cos\theta_{q},\chi),
\end{equation}
where $J=1$ for $\lambda_{V}=\pm1,0$, $J=0$ for $\lambda_{V}=t$ and
correspondingly for the primed quantities. It turns out that $J=J'$ in the
decay $H\to Z(\to e^{+}e^{-})+Z^{\ast}(\to \tau^{+}\tau^{-})$ as long as one
is not analyzing $\tau$-polarization effects, i.e.\ there are no
spin-0 -- spin-1 interference effects in this decay.

The evaluation of the helicity components of the $H \to VV^{\ast}$ transition
amplitudes $H_{\lambda_{V},\lambda_{V^{\ast}}}$ is given in Appendix~A while
the evaluation of the helicity components of the lepton tensors
$L_{\lambda_{V}\,\lambda'_{V}}^{(p)}(\cos\theta_{p})$ and
$L^{(q)}_{\lambda_{V^{\ast}}\,\lambda'_{V^{\ast}}}(\cos\theta_{q}, \chi)$ are
given in Appendix~B (neutral-current case) and~C (charged-current case).

Up to this point we have allowed for a general structure of the $(HVV)$
coupling. In the following we shall specify to the SM coupling with
$H_{\alpha\alpha'}=g_{\alpha\alpha'}$. 

\section{The four-body decay  $H\to Z(\to \ell^{+}\ell^{-})
+Z^{\ast}(\to \tau^{+}\tau^{-})$}
In this section we write down the three-fold angular decay distribution of the
decay $H\to Z(\to \ell^{+}\ell^{-})+Z^{\ast}(\to \tau^{+}\tau^{-})$ involving
two different pairs of leptons, i.e. we assume $\ell\ne\tau$. The
corresponding decay
$H\to Z(\to \ell^{+} \ell^{-})+Z^{\ast}(\to \ell^{+} \ell^{-})$ involving two
pairs of identical leptons (with and without lepton-mass effects) is more
difficult to analyze due to the presence of nonfactorizing interference 
contributions. These identical-particle effects will be treated in a separate
paper~\cite{Groote:2015}.

\subsection{Three-fold angular decay  distribution for the \\
four-body decay $H \to Z(\to \ell^{+}\ell^{-})+Z^{\ast}(\to \tau^{+}\tau^{-})$}
We begin our discussion by presenting an explicit form of the three-fold
angular decay distribution given by Eq.~(\ref{angdis2}). The relevant helicity
components of the on-shell and off-shell lepton tensors are listed in
Appendix~B while the helicity components of the $H\to Z Z^{\ast}$ transition
amplitude can be found in Appendix~A. The polar angles $\theta_{p}$ and
$\theta_{q}$ are defined in the respective lepton pair center-of-mass systems
as shown in Fig.~\ref{planeszz}. The azimuthal angle $\chi$ describes the
relative orientation of the two decay planes. We split the decay distribution
into a helicity-nonflip and helicity-flip part, 
\begin{eqnarray}\label{Zjoint3}
\lefteqn{(2p^{2}2q^{2})^{-1}\,W^{Z}_{nf}(\theta_{p},\theta_{q},\chi)
  \ =\ (\rho_{++}+\rho_{--})}\nn&&\qquad\strut\times
  \Big(\tfrac14(1+\cos^{2}\theta_{p})(v_{\ell}^{2}+a_{\ell}^{2}v_{p}^{2})
    (1+\cos^{2}\theta_{q})(v_{\ell}^{2}+a_{\ell}^{2}v_{q}^{2})
    +4\cos\theta_{p}\cos\theta_{q}v_{\ell}^{2}a_{\ell}^{2}v_{p}v_{q}
  \Big)\kern-2pt\nn&&\strut
  +\rho_{00}\sin^{2}\theta_{p}(v_{\ell}^{2}+a_{\ell}^{2}v_{p}^{2}) 
    \sin^{2}\theta_{q}(v_{\ell}^{2}+a_{\ell}^{2}v_{q}^{2})
  +(\rho_{++}-\rho_{--})\nn&&\qquad\strut\times
  \left((1+\cos^{2}\theta_{p})(v_{\ell}^{2}+a_{\ell}^{2}v_{p}^{2})
  \cos\theta_{q}v_{q}+\cos\theta_{p}v_{p}(1+\cos^{2}\theta_{q})
  (v_{\ell}^{2}+a_{\ell}^{2}v_{q}^{2})\right)v_{\ell}a_{\ell}\nn&&\strut
  +(\rho_{+0}+\rho_{-0})\sin\theta_{p}\sin\theta_{q}
    \Big(4v_{\ell}^{2}a_{\ell}^{2}v_{p}v_{q}
    +\cos\theta_{p}(v_{\ell}^{2}+a_{\ell}^{2}v_{p}^{2})\cos\theta_{q}
    (v_{\ell}^{2}+a_{\ell}^{2}v_{q}^{2})\Big)\cos\chi\nn&&\strut
  +2(\rho_{+0}-\rho_{-0})\sin\theta_{p}\sin\theta_{q}v_{\ell}a_{\ell}
  \left(\cos\theta_{p}(v_{\ell}^{2}+a_{\ell}^{2}v_{p}^{2})v_{q}
  +\cos\theta_{q}(v_{\ell}^{2}+a_{\ell}^{2}v_{q}^{2})v_{p}\right)
  \cos\chi\nn&&\strut
  +\tfrac12\rho_{+-}
  \sin^{2}\theta_{p}(v_{\ell}^{2}+a_{\ell}^{2}v_{p}^{2})
  \sin^{2}\theta_{q}(v_{\ell}^{2}+a_{\ell}^{2}v_{q}^{2})\cos2\chi
\end{eqnarray}
and
\begin{eqnarray}\label{Zjoint4}
\lefteqn{(2p^{2}2q^{2})^{-1}\,W^{Z}_{hf}(\theta_{p},\theta_{q},\chi)  
\ =\ \frac{4m_{\tau}^{2}}{q^{2}}\Big\{(v_{\ell}^{2}+a_{\ell}^{2}v_{p}^{2})
  \times\strut}\nn&&
\Big(\tfrac14(\rho_{++}+\rho_{--})(1+\cos^{2}\theta_{p})
    \sin^{2}\theta_{q}v_{\ell}^{2}
  +\rho_{00}\sin^{2}\theta_{p}\cos^{2}\theta_{q}v_{\ell}^{2}
  +\rho_{S}\,F^{2}_{S}\sin^{2}\theta_{p}a_{\ell}^{2}\nn&&\strut
  -\tfrac14(\rho_{+0}+\rho_{-0})\sin2\theta_{p}\sin2\theta_{q}
  v_{\ell}^{2}\cos\chi
  -\tfrac12\rho_{+-}\sin^{2}\theta_{p}\sin^{2}\theta_{q}v_{\ell}^{2}
  \cos2\chi\Big)\nn&&\strut
  +(\rho_{++}-\rho_{--})\cos\theta_{p}v_{p}
  \sin^{2}\theta_{q}v_{\ell}^{3}a_{\ell}
  -(\rho_{+0}-\rho_{-0})\sin\theta_{p}v_{p}\sin2\theta_{q}
  v_{\ell}^3a_{\ell}\cos\chi\Big\},\qquad
\end{eqnarray}
where $v_{p}^{2}=1-4m_{\ell p}^{2}/p^{2}$ and
$v_{q}^{2}=1-4m_{\ell q}^{2}/q^{2}$. For symmetry reasons and for later
applications in the off-shell -- off-shell case we have written $p^{2}$ for
$m_{Z}^{2}$ and $v_{p}^{2}=1-4m_{\ell p}^{2}/p^{2}$ for $v_{p}^{2}=1$ on the
on-shell side. The double spin-density matrix elements $\rho_{mm'}$ are 
bilinear forms of the helicity amplitudes describing the $H\to ZZ^{\ast}$ 
transitions. They are defined in Appendix~A.

In Eqs.~(\ref{Zjoint3}) and~(\ref{Zjoint4}) we have also included the
contributions from the parity-violating terms proportional to
$(\rho_{++}-\rho_{--})$ and $(\rho_{+0}-\rho_{-0})$. These
coefficient functions are not populated by the parity-conserving SM $(HVV)$
coupling. In Appendix~A we briefly discuss the contribution of a
parity-violating non-SM coupling proportional to
$\epsilon^{\mu\nu\rho\sigma}p_{\rho}q_{\sigma}$ which would populate the
$(\rho_{++}-\rho_{--})$ and $(\rho_{+0}-\rho_{-0})$ coefficient
functions~\cite{Modak:2013sb,Bhattacherjee:2015xra,Zagoskin:2015sca}.   

We add the flip and non-flip contributions and expand the result in terms of
the Legendre polynomials $P_{1}(\cos\theta)=\cos\theta$ and
$P_{2}(\cos\theta)=\frac12(3\cos^2\theta-1)$. The result is written in the 
form
\begin{equation}\label{eight}
(2p^{2}2q^{2})^{-1}\,W^{Z}(\theta_{p},\theta_{q},\chi)
  =\frac49\sum_{i=0}^{7}{\cal F}^{Z}_{i}h_{i}(\theta_{p},\theta_{q},\chi)
  =\frac49\sum_{i=0}^{7}\Big(f^{Z}_{i}+\eps g^{Z}_{i}\Big)
  h_{i}(\theta_{p},\theta_{q},\chi),
\end{equation}
where in the second equation of~(\ref{eight}) we have split the coefficient
function ${\cal F}^{Z}_{i}$ into its helicity-flip and helicity-nonflip
part using the notation $\eps=m_{\tau}^{2}/q^{2}$.

The coefficient functions $f^{Z}_{i}$ and $g^{Z}_{i}$ and their associated
angular factors $h_{i}(\theta_{p},\theta_{q},\chi)$ are listed in
Table~\ref{tseven} where we use the abbreviation
$C^{(i)}_{ew}=v_{\ell}^{2}+a_{\ell}^{2}v_{i}^{2}$ with $i=p,q$. In addition
we use a short-hand notation for the double density matrix elements, namely
$\rho_{U}=\rho_{++}+\rho_{--}$, $\rho_{L}=\rho_{00}$,
$\rho_{U+L}=\rho_{U}+\rho_{L}$ and $\rho_{S}=\rho_{tt}$. Note that we have
dropped the contributions of the parity-violating terms proportional to
$(\rho_{++}-\rho_{--})$ and $(\rho_{+0}-\rho_{-0})$ in Table~\ref{tseven}
which are not populated by the parity-conserving SM $(HVV)$ coupling. 

The dominant flip contributions proportional to $a^{2}_{\ell}$ are contained 
in $g^{Z}_0$ and $g^{Z}_2$. Compared to $a_{\ell}$ the leptonic vector 
coupling $v_\ell=-1+\sin^{2}\theta_{W}$ is much suppressed. This is
different in the quark--antiquark case treated in
Refs.~\cite{Groote:2012xr,Groote:2013xt} where the electroweak vector and
axial vector couplings to the quark pairs are comparable in size. As a result
the pattern of the helicity-flip contributions in the quark pair production
case is quite different from the lepton-pair production
case~\cite{Groote:2012xr,Groote:2013xt}.
\begin{table}[t]
\begin{center}\begin{tabular}{llll}
\hline\noalign{\vskip 2mm}
$i$\hspace*{0.5cm} & $f^{Z}_{i}$\hspace*{2.5cm} & $g^{Z}_{i}$\hspace*{3.5cm} & 
    $h_i(\theta_{p},\theta_{q},\chi)$ 
\\ \noalign{\vskip 2mm}\hline\noalign{\vskip 2mm}
0   & $C^{(p)}_{ew}C^{(q)}_{ew}\,\rho_{U+L}$
& $2C^{(p)}_{ew}(v_{\ell}^{2}\rho_{U+L}
   +3a_{\ell}^{2}F_{S}^{2}\rho_{S})$
    & $1$ 
\\ \noalign{\vskip 1mm}
1   & $\frac 12 C^{(p)}_{ew}C^{(q)}_{ew}(\rho_{U}-2\rho_{L})$ & 
$-2C^{(p)}_{ew}v_{\ell}^{2}(\rho_{U}-2\rho_{L})$   
& $P_{2}(\cos\theta_{q})$     
\\ \noalign{\vskip 1mm}
2   & $\frac 12C^{(p)}_{ew}C^{(q)}_{ew} (\rho_{U}-2\rho_{L})$   & 
$C^{(p)}_{ew}\,\Big(v_{\ell}^{2} 
(\rho_{U}-2\rho_{L})-6a_{\ell}^{2}F_{S}^{2}\rho_{S}$\Big)  
& $P_{2}(\cos\theta_{p})$ 
\\ \noalign{\vskip 1mm}
3   & $\frac 14 C^{(p)}_{ew}C^{(q)}_{ew}(\rho_{U}+4\rho_{L})$  & 
$-C^{(p)}_{ew}v_{\ell}^{2} 
(\rho_{U}+4\rho_{L})$  &   
      $P_{2}(\cos\theta_{p})P_{2}(\cos\theta_{q})$   
\\ \noalign{\vskip 1mm}
4   & $  9 v_{\ell}^{2}a_{\ell}^{2}v_{p}v_{q}\rho_{U}$  & $0$  &  
$\cos\theta_{p}\cos\theta_{q}$
\\ \noalign{\vskip 1mm}
5   & $9v_{\ell}^{2}a_{\ell}^{2}v_{p}v_{q}
(\rho_{+0}+\rho_{-0})$    & 
$0$  &  
      $\sin\theta_{p}\sin\theta_{q}\cos\chi$ 
\\ \noalign{\vskip 1mm}
6   & $\frac{9}{16}C^{(p)}_{ew}C^{(q)}_{ew}(\rho_{+0}+\rho_{-0})$    & 
$-\frac{9}{4}C^{(p)}_{ew}v_{\ell}^2(\rho_{+0}+\rho_{-0})$  &  
      $\sin2\theta_{p}\sin2\theta_{q}\cos\chi$ 
\\ \noalign{\vskip 1mm}
7  & $\tfrac{9}{8}C^{(p)}_{ew}C^{(q)}_{ew}
\rho_{+-}$    & $-\frac 92 C^{(p)}_{ew}v_{\ell}^{2}
\rho_{+-}$  &  
      $\sin^{2}\theta_{p}\sin^{2}\theta_{q}\cos2\chi$ 
\\[1.5ex]
\hline
    \end{tabular}
  \end{center}
\caption{\label{tseven}Coefficient functions appearing in the three-fold
angular decay distribution of the decay
$H \to Z^{*0}(\to \ell^{+}\ell^{-})+Z^{\ast}(\to \tau^{+}\tau^{-})$}
\end{table}

In order to save space we have not expanded the angular factor 
$\sin^{2}\theta_{p}\sin^{2}\theta_{q}$ in the last row of Table~\ref{tseven}.
The relevant expansion would be given by
\begin{equation}
\sin^{2}\theta_{p}\sin^{2}\theta_{q}=\frac 49\,\bigg(1-P_{2}(\cos\theta_{p})
-P_{2}(\cos\theta_{q})+P_{2}(\cos\theta_{p})P_{2}(\cos\theta_{q})\bigg).
\end{equation}

We then define a normalized decay distribution
\begin{equation}\label{tildeW}
\widetilde W^{Z}(\theta_{p},\theta_{q},\chi)
  =\frac{W^{Z}(\theta_{p},\theta_{q},\chi)}{\int W^{Z}(\theta'_{p},
  \theta'_{q},\chi')d\cos\theta'_{p}\,d\cos\theta'_{q}\,d\chi'}
  =\frac{1}{8\pi}\Big(1+\sum_{i=1}^{7}{\cal\widetilde F}^{Z}_{i}
  h_{i}(\theta_{p},\theta_{q},\chi)\Big),
\end{equation}
where ${\cal\widetilde F}^{Z}_i={\cal F}^{Z}_i/{\cal F}^{Z}_0$
(and $\tilde f^{Z}_{i}=f^{Z}_{i}/{\cal F}^{Z}_{0}$,
$\tilde g^{Z}_{i}=g^{Z}_{i}/{\cal F}^{Z}_0$) and where
\begin{equation}\label{norm}
{\cal F}^{Z}_0=f^{Z}_{0}+\eps g^{Z}_{0}
  =C^{(p)}_{ew}C^{(q)}_{ew}\rho_{U+L}+2\eps C^{(p)}_{ew}
  (v_{\ell}^{2}\rho_{U+L}+3a_{\ell}^{2}F_{S}^{2}\rho_{S}).
\end{equation}
The normalized angular decay distribution 
$\widetilde W^{Z}(\theta_{p},\theta_{q},\chi)$ obviously integrates 
to $1$, i.e.\
\begin{equation}
\int \widetilde W^Z(\theta_{p},\theta_{q},\chi)\,
d\cos\theta_{p}\,d\cos\theta_{q}\,d\chi\,=\,1.
\end{equation}

Before we start discussing our numerical results we want to specify our mass,
width and coupling input parameters. We use the central value of the Higgs 
mass $m_{H}=125.09(24)\GeV$ from the combined ATLAS and CMS
measurement~\cite{Aad:2015zhl}. For the remaining parameters we use the
central values from the PDG~\cite{Agashe:2014kda} given by 
\begin{eqnarray}
m_{W}=80.385(15)\GeV,&&\Gamma_{W}=2.085(42)\GeV,\nn
m_{Z}=91.1876(21)\GeV,&&\Gamma_{Z}=2.4952(23)\GeV,\nn
m_{\tau}=1.77682(16)\GeV,&&\nn
\sin^{2}\theta_{W}=0.23126(5),&&G_F=1.1663787(6)\times 10^{-5}\GeV^{-2}.\qquad
\end{eqnarray}
Our formulas are written in terms of the dimensionless coupling constant 
$g^{2}$ which is related to $G_{F}$ by $g^{2}=8m_{W}^{2}G_{F}/\sqrt{2}$.
For practical numerical purposes we choose $m_{\ell p}=m_e$ (or
$m_{\ell p}=0$) on the $p$ side. On the off-shell $q$ side we write 
$m_{\ell q}=m_{\ell}$ which can take the values $m_{\ell}=m_{\tau}$ or
$m_{\ell}=m_{e,\mu}$.

\begin{table}
\begin{center}\begin{tabular}{lllll}\hline\noalign{\vskip 2mm}
$i$&${\cal\widetilde F}^{Z}_i$ ($m_{\ell}=0$)
&${\cal\widetilde F}^{Z}_i$ ($m_{\ell}=m_{\tau}$)
&$\langle{\cal\widetilde F}^{Z}_i\rangle$ ($m_{\ell}=0$)
&$\langle{\cal\widetilde F}^{Z}_i\rangle$ ($m_{\ell}=m_{\tau}$)
  \\\noalign{\vskip 2mm}\hline\noalign{\vskip 2mm}
$1$&$-0.9115$&$-0.6257$&$-0.3916$&$-0.3491$\\\noalign{\vskip 2mm}
$2$&$-0.9115$&$-0.9391$&$-0.3916$&$-0.3908$\\\noalign{\vskip 1mm}
$3$&$+0.9557$&$+0.6561$&$+0.6958$&$+0.6537$\\\noalign{\vskip 1mm}
$4$&$+0.0030$&$+0.0023$&$+0.0203$&$+0.0206$\\\noalign{\vskip 1mm}
$5$&$+0.0167$&$+0.0132$&$+0.0319$&$+0.0319$\\\noalign{\vskip 1mm}
$6$&$+0.1875$&$+0.1287$&$+0.3589$&$+0.3528$\\\noalign{\vskip 1mm}
$7$&$+0.0332$&$+0.0228$&$+0.2281$&$+0.2284$\\\noalign{\vskip 2mm}
\hline
\end{tabular}\end{center}
\caption{\label{normz}Numerical results for the normalized coefficient
functions ${\cal\widetilde F}^{Z}_{i}(q^{2})$ at $q^{2}=50\GeV^{2}$ and the
average of ${\cal\widetilde F}^{Z}_{i}(q^{2})$ over
$q^{2}\in[4m_{\ell}^{2},(m_{H}-m_{Z})^2]$}
\end{table}

In Table~\ref{normz} we present numerical results for the normalized
coefficient functions ${\cal\widetilde F}^{Z}_{i}(q^{2})$ and their averages.
In  columns~2 and~3 we list the values of ${\cal\widetilde F}^{Z}_{i}(q^{2})$
for $q^{2}=50\GeV^{2}$ with zero and nonzero lepton masses. In order to
avoid possible contamination from contributions of the $\psi$ and $\Upsilon$ 
families we have chosen a $q^{2}$ value in between these two families, namely
$q^{2}=50\GeV^{2}$. This $q^{2}$ value is small enough to highlight the 
helicity-flip and lepton-mass effects in the vicinity of the threshold. On the
other hand, this value of $q^{2}$ is far away enough from the threshold region 
where one would have to deal with the Coulomb singularity. We mention that the
contribution of the $\psi$ and $\Upsilon$ families to the $q^{2}$ spectrum
have been investigated in Ref.~\cite{Gonzalez-Alonso:2014rla}. These
contributions have been found to be small.

Concerning the $q^{2}=50\GeV^{2}$ values for the normalized coefficient
functions, lepton-mass effects amount to $-31\,\%$ for the functions
${\cal\widetilde F}^{Z}_{1,3,6,7}$, $-21\,\%$ for the functions
${\cal\widetilde F}^{Z}_{4,5}$, and $+3\,\%$ for the function
${\cal\widetilde F}^{Z}_{2}$. The normalized coefficient functions
${\cal\widetilde F}^{Z}_{6,7}$ are quite small to start with. We mention that
$\tau$-lepton mass effects are even larger for smaller values of $q^{2}$.

In columns~4 and~5 of Table~\ref{normz} we also present average values
$\langle{\cal\widetilde F}^{Z}_{i}\rangle$  of the coefficient functions
again for zero and nonzero lepton masses where the average is taken with
regard to $q^{2}$. In order to do the requisite $q^{2}$ integrations one 
needs to include the relevant $q^{2}$-dependent integration measure defined by
the differential $q^{2}$ distribution. Inserting the necessary coupling and
phase-space factors one obtains
\begin{equation}
\frac{d\Gamma^{Z}}{dq^{2}\,d\cos\theta_{p}\,d\cos\theta_{q}\,d\chi}
  =\frac{B_{Z\ell\ell}}{C_{ew}^{(p)}}\,\frac{C^{Z}(q^{2})}{8\pi}\times
  \frac94W^{Z}(q^{2},\theta_{p},\theta_{q},\chi),
\end{equation}
where
\begin{equation}
C^{Z}(q^{2})=\frac{g^{4}}{\cos^{4}\theta_{W}}
\frac{1}{4\cdot1536\pi^{3}}
\frac{|\vec p_{V}(m_{Z}^{2},q^{2})|v_{q}}{m_{H}^{2}}
\frac{1}{(q^{2}-m_{Z}^{2})^{2}+m_{Z}^{2}\Gamma_{Z}^{2}}\,.
\end{equation}
$B_{Z\ell\ell}$ is the branching ratio of the decay
$Z\to\ell^{+}\ell^{-}$, i.e.\
$B_{Z\ell\ell}=\Gamma(Z \to \ell^{+}\ell^{-})/\Gamma(Z)$,
where the rate for the decay $Z\to \ell^{+}\ell^{-}$ ($m_{\ell p}=0$) reads
\begin{equation}\label{gamZell}
\Gamma_{Z\ell\ell}=\Gamma(Z\to \ell^{+}\ell^{-})
  =\frac{g^{2}}{\cos^{2}\theta_{W}}
  \frac{1}{192\pi}m_{Z}\,(v_{\ell}^{2}+a_{\ell}^{2}).
\end{equation}
The magnitude of the momentum of the gauge bosons is given by
\begin{equation}
|\vec p_{V}(p^{2},q^{2})|=\frac1{2m_H}\sqrt{\lambda(m_H^2,p^2,q^2)},
\end{equation}
where $\lambda(a,b,c)=a^2+b^2+c^2-2ab-2ac-2bc$ is K\"all\'en's function. 
We then define partial differential rates according to
\begin{equation}\label{partialz}
\frac{d\,\Gamma^{Z}_{i}}{dq^{2}}
  =2p^{2}\,2q^{2}\frac{B_{Z\ell\ell}C^{Z}(q^{2})}{C^{(p)}_{ew}}
  {\cal F}^{Z}_{i}(q^{2}).
\end{equation}
The factors $p^{2}=m_{Z}^{2}$ and $q^{2}$ are picked up when doing the 
integrations over the $Z\to\ell^{+}\ell^{-}$ and $Z^{\ast}\to\ell^{+}\ell^{-}$
phase spaces. The factor $q^{2}$ is of crucial importance to cancel the
$1/q^2$ singularity in the double spin-density matrix elements $\rho_{00}$,
$\rho_{0t}$ and $\rho_{tt}$ at the lower end of the $q^2$ spectrum. The
average values of the coefficient functions
$\langle{\cal\widetilde F}^{Z}_{i}\rangle$ can be calculated from the formula
\begin{equation}\label{averagez}
\langle{\cal\widetilde F}^{Z}_{i}\rangle
=\frac{\int dq^{2}2p^{2}2q^{2}B_{Z\ell\ell}C^{Z}(q^{2})
  {\cal F}^{Z}_{i}(q^{2})/C^{(p)}_{ew}}{\int dq^{2}2p^{2}2q^{2}
  B_{Z\ell\ell}C^{Z}(q^{2}){\cal F}^{Z}_{0}(q^{2})/C^{(p)}_{ew}}
=\frac{\Gamma^{Z}_{i}}{\Gamma^{Z}}.
\end{equation}
The integration has to be done in the limits
$4m_{\ell}^{2}\le q^{2}\le (m_{H}-m_{Z})^{2}$. The denominator of
Eq.~(\ref{averagez}) is nothing but the total rate $\Gamma^{Z}$ including
lepton mass effects. When calculating the average values according to
Eq.~(\ref{averagez}) one can disregard all constant factors in the weight
function $C^{Z}(q^{2})$.

Numerically, one finds that the averaged coefficient function
$\langle{\cal\widetilde F}^{Z}_{3}\rangle$ is by far the largest one.
Lepton-mass effects are largest for
$\langle{\cal\widetilde F}^{Z}_{1,3}\rangle$ and amount to $-10.8\,\%$ and
$-6.1\,\%$.

Using the average values $\langle{\cal\widetilde F}^{Z}_{i}\rangle$, the
$q^{2}$-integrated angular decay distribution can be written as
\begin{equation}
\frac{1}{\Gamma^{Z}}
  \frac{d\,\Gamma^{Z}}{d\cos\theta_{p}\,d\cos\theta_{q}\,d\chi}
  =\frac{1}{8\pi}\Big(\,1+\sum_{i=1}^{7}\langle{\cal\widetilde F}^{Z}_{i}
  \rangle\,h_{i}(\theta_{p},\theta_{q},\chi)\Big).
\end{equation}

One can make contact with the work of Ref.~\cite{Keung:1984hn} by taking the
zero-mass limit $m_{\ell q}\to 0$ of Eq.~(\ref{partialz}) (for $i=0$), 
neglecting the $Z$ width and omitting the factor $B_{Z\ell\ell}$. In fact,
using the notation $\hat m_{Z}=m_{Z}/m_{H}$ and $\hat q^{2}=q^{2}/m_{H}^{2}$
one obtains
\begin{equation}\label{keung}
\frac{d\Gamma^{Z}}{d\hat q^{2}}=\frac{g^{4}}{\cos^{4}\theta_{W}}
\frac{m_{H}C^{(q)}_{ew}}{4\cdot 3072\pi^{3}}
\lambda^{1/2}(1,\hat m_{Z}^{2},\hat q^{2})\,\,
\frac{\hat q^{4}+\hat q^{2}(10\hat m_{Z}^{2} - 2)+(1-\hat m_{Z}^{2})^{2}}
{(\hat q^{2}-\hat m_{Z}^{2})^{2}}
\end{equation}
in agreement with Refs.~\cite{Keung:1984hn,Djouadi:2005gi}. At $\hat q^{2}=0$
one has
\begin{equation}\label{keungnum}
\frac{d\Gamma^{Z}}{d\hat q^{2}}(\hat q^{2}=0)=
\frac{g^{4}}{\cos^{4}\theta_{W}}
\frac{m_{H}C^{(q)}_{ew}}{4\cdot 3072\pi^{3}}
\frac{(1-\hat m_{Z}^{2})^{3}}{\hat m_{Z}^{4}}=4.02\cdot 10^{-2}\MeV.
\end{equation}
Integrating Eq.~(\ref{keung}) within the limits
$0\le \hat q^{2}\le(1-\hat m_{Z})^{2}$  
one obtains the well-known expression~\cite{Keung:1984hn,Djouadi:2005gi}
\begin{equation}
\Gamma(H \to Z\,\ell^{+}\ell^{-})=\frac{g^{4}}{\cos^{4}\theta_{W}}
\frac{m_{H}}{4\cdot3072\pi^{3}}
F(\hat m_{Z})
\end{equation}
with
\begin{eqnarray}
F(\hat m_{Z})&=&\frac{3(1-8\hat m_{Z}^{2}+20\hat m_{Z}^{4})}
{(4\hat m_{Z}^{2}-1)^{1/2}}\arccos\bigg(\frac{3\hat m_{Z}^{2}-1}
{2\hat m_{Z}^{3}}\bigg)\nn&&\strut
-(1-\hat m_{Z}^{2})\Big(\frac{47}{2}\hat m_{Z}^{2}-\frac{13}{2}
+\frac{1}{\hat m_{Z}^{2}}\Big) -3(1-6\hat m_{Z}^{2}+4\hat m_{Z}^{4})
\ln\hat m_{Z}.
\end{eqnarray}

\subsection{Single-angle decay distributions\label{sec32}}
\begin{figure}\begin{center}
\epsfig{figure=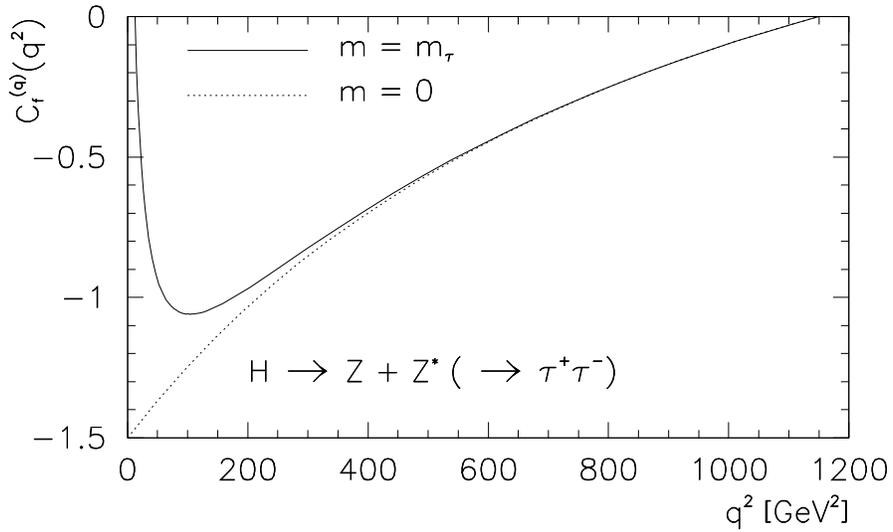, scale=0.7}
\caption{\label{dgamhzc}$q^{2}$ distribution of the convexity parameter
for $m_{\ell}=0$ (dotted line) and $m_{\ell}=m_{\tau}$ (solid line) for
the process $H\to Z+Z^{\ast}(\to\ell^{+}\ell^{-})$}
\end{center}\end{figure}

Integrating Eq.~(\ref{tildeW}) over $\cos\theta_{p}$ and $\chi$ and using
Table~\ref{tseven}, one obtains
\begin{equation}\label{theta1dis}
\widetilde W^Z(q^{2},\theta_{q})
  =\frac12\left(1+{\cal\widetilde F}^{Z}_{1}P_{2}(\cos\theta_{q})\right).
\end{equation}
We define a convexity parameter $C_{f}^{(q)}(q^{2})$ as the second derivative
of Eq.~(\ref{theta1dis}) with respect to $\cos\theta_{q}$ or, equivalently, as
two times the coefficient of the $\cos^{2}\theta_{q}$ term in
Eq.~(\ref{theta1dis}). One obtains
\begin{equation}\label{convex}
C_{f}^{(q)}(q^{2})=\frac 32\,{\cal\widetilde F}^{Z}_{1}=\frac 32\,
  \frac{f^{Z}_{1}+\eps g^{Z}_{1}}{f^{Z}_{0}+\eps g^{Z}_{0}}
  =\frac 34\,\frac{\rho_{U}-2\rho_{L}}{\rho_{U+L}}\,
\frac{(q^{2}-4m_{\ell}^{2})(v_{\ell}^{2}+a_{\ell}^{2})}{C^{(q)}_{ew}q^{2}
  +2m_{\ell}^{2}(v_{\ell}^{2}+3a_{\ell}^{2}F_{S}^{2}\rho_{S}/\rho_{U+L})}.
\end{equation}
In Fig.~\ref{dgamhzc} we show a plot of the $q^{2}$ distribution of the
convexity parameter for both the $m_{\ell}=0$ and $m_{\ell}=m_{\tau}$ cases.
Due to the overall factor $(q^{2}-4m_{\ell}^{2})$ one has
$C_{f}^{(q)}(q^{2}) \to 0$ at threshold (maximal recoil)
$q^{2}=4m_{\ell}^{2}$, i.e.\ the $\cos^{2}\theta_{q}$ distribution is flat at
threshold. This is clearly visible in the $\ell=\tau$ case in
Fig.~\ref{dgamhzc}. For $\ell=e,\mu$ the vanishing of the convexity parameter
$C_{f}^{(q)}(q^{2})$ at threshold is not discernible at the scale of
Fig.~\ref{dgamhzc}. Instead $C_{f}^{(q)}(q^{2})\to-3/2$ at threshold as
$m_{\ell}^{2} \to 0$ due to the limiting behaviour of the two factors in
Eq.~(\ref{convex}). The first factor goes to $-3/2$ because of the dominance
of $\rho_{L}$ and the second factor goes to $1$ in this limit. A closer look
at the second factor in Eq.~(\ref{convex}) reveals that it shows a steplike
behaviour at threshold for $m_{\ell}\to 0$ jumping from $0$ to $1$. In fact,
setting $F_{S}^{2}\,\rho_{S}/\rho_{U+L}\sim\,1$ in the second factor in
Eq.~(\ref{convex}) one obtains the limiting form
\begin{equation}\label{limbeh}
\frac{(q^{2}-4m_{\ell}^{2})(v_{\ell}^{2}+a_{\ell}^{2})}{C^{(q)}_{ew}q^{2}
  +2m_{\ell}^{2}(v_{\ell}^{2}+3a_{\ell}^{2}F_{S}^{2}\rho_{S}/\rho_{U+L})}
  \quad\to\quad (q^{2}-4m_{\ell}^{2})/(q^{2}+2m_{\ell}^{2})\,. 
\end{equation}
Equation~(\ref{limbeh}) shows the advertised steplike behaviour as 
$m_{\ell}^{2} \to 0$. Even for the muon the deviation from $1$ at 
$q^{2}=10\GeV^{2}$ is a tiny one ($0.67\,\%$). At the other end of the
spectrum at minimal (zero) recoil where $q^{2}=(m_{H}-m_{Z})^{2}$ the
convexity parameter goes to zero in both cases since
$\rho_{U}-2\rho_{L}\sim\,|\vec q|^{2}$ and $|\vec q|=0$ at zero recoil. 

\begin{figure}\begin{center}
\epsfig{figure=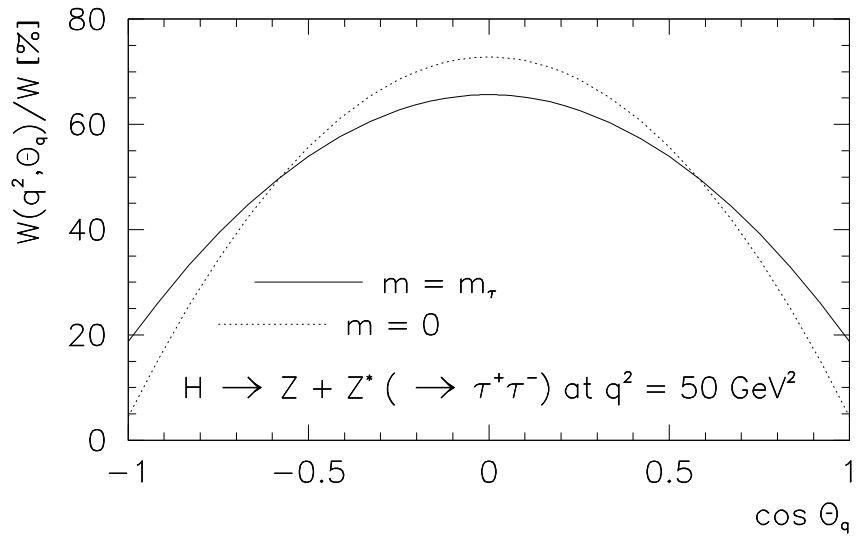, scale=0.7}
\caption{\label{wthq}$\cos\theta_{q}$ dependence of the normalized decay
distribution $\widetilde W^{Z}(q^{2},\theta_q)$ for $q^{2}=50\GeV^{2}$ in case
of $m_{\ell}=0$ (dotted line) and $m_{\ell}=m_{\tau}$ (solid line)}
\end{center}\end{figure}

Figure~\ref{dgamhzc} shows that the convexity parameter $C^{(q)}_{f}$ is
negative, i.e.\ the polar angle distribution is described by a downward-open
parabola which has a maximum at $\cos\theta_{q}=0$ with the maximal value 
$\widetilde W^{Z}(q^{2},\theta_{q}=\pi/2)=\frac12(1-C_{f}^{(q)}(q^{2})/3)$. 
In Fig.~\ref{wthq} we show a plot of the $\cos\theta_{q}$ distribution for
$q^{2}=50\GeV^{2}$. The $\cos\theta_{q}$ distribution is symmetric in
$\cos\theta_{q}$ due to the absence of a term linear in $\cos\theta_{q}$ in
Eq.(\ref{theta1dis}), i.e.\ the distribution is forward--backward symmetric.
As expected from Eq.~(\ref{limbeh}) the $m_{\ell}=m_{\tau}$ curve is
considerably flatter than the $m_{\ell}=0$ curve. At $\cos\theta_{q}=\pm 1$
the $m_{\ell}=0$ curve is close to zero since $P_{2}(\cos\theta_{q})=1$ at
these points and ${\cal\widetilde F}^{Z}_{2}=f_{1}^{Z}/f_{0}^{Z}\approx -1$
due to the dominance of the longitudinal contribution $\rho_{L}$. 

\begin{figure}\begin{center}
\epsfig{figure=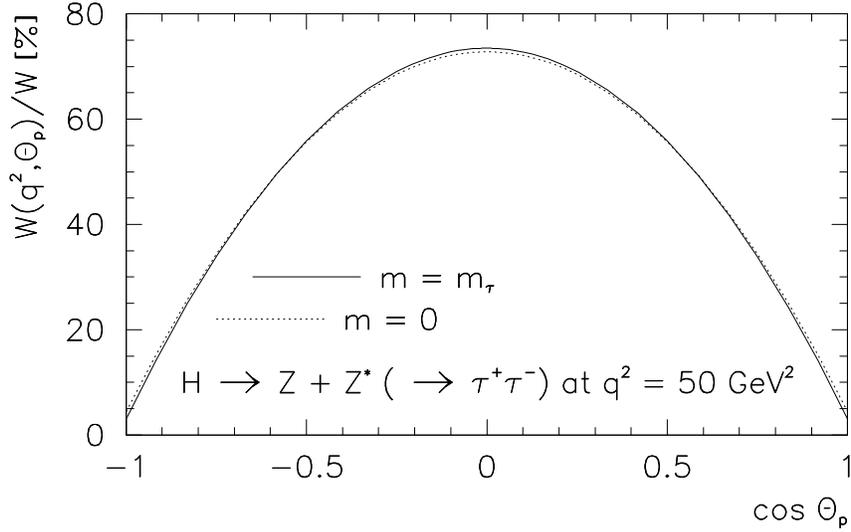, scale=0.7}
\caption{\label{wthp}$\cos\theta_{p}$ dependence of the normalized decay
distribution $\widetilde W^{Z}(q^{2},\theta_p)$ for $q^{2}=50\GeV^{2}$ in case
of $m_{\ell}=0$ (dotted line) and $m_{\ell}=m_{\tau}$ (solid line)}
\end{center}\end{figure}

Next we discuss the single-angle $\cos\theta_{p}$ distribution. From
Table~\ref{tseven} one reads off
\begin{equation}\label{theta2dis}
\widetilde W^{Z}(q^{2},\theta_{p})=
\frac12\left(1+{\cal\widetilde F}^{Z}_{2}P_{2}(\cos\theta_{p})\right).
\end{equation}
The corresponding convexity factor is now given by
\begin{equation}
C_{f}^{(p)}(q^{2})=\frac 32\,{\cal\widetilde F}^{Z}_{2}
  =\frac 32\,\frac{f^{Z}_{2}+\eps g^{Z}_{2}}{f^{Z}_{0}+\eps g^{Z}_{0}}
  =\frac 34\,\frac{\rho_{U}-2\rho_{L}}{\rho_{U+L}}\,
  \frac{C^{(q)}_{ew}q^{2}+2m_{\ell}^{2}(v_{\ell}^{2}
  -6a_{\ell}^{2}F_{S}^{2}\rho_{S}/\rho_{U-2L})}{C^{(q)}_{ew}q^{2}
  +2m_{\ell}^{2}(v_{\ell}^{2}
  +3a_{\ell}^{2}F_{S}^{2}\rho_{S}/\rho_{U+L})}.
\end{equation}
The threshold value of the convexity parameter can now be seen to be given 
by $C^{(p)}_{f}(q^{2})=-3/2$ in both the $m_{\ell}=0$ and $m_{\ell}=m_{\tau}$
cases. We do not provide a plot of the convexity parameter $C_{f}^{(p)}(q^{2})$
because lepton-mass effects are small even close to threshold. In
Fig.~\ref{wthp} we plot the $\cos\theta_{p}$ dependence of the normalized
single-angle distribution again for $q^{2}=50\GeV^{2}$. There is practically
no lepton-mass dependence in the $\cos\theta_{p}$ distribution. Since
${\cal\widetilde F}^{Z}_{2}={\cal\widetilde F}^{Z}_{1}$ in the zero mass case,
the zero lepton-mass distributions in Figs.~\ref{wthq} and~\ref{wthp} are
identical to each other.

\begin{figure}\begin{center}
\epsfig{figure=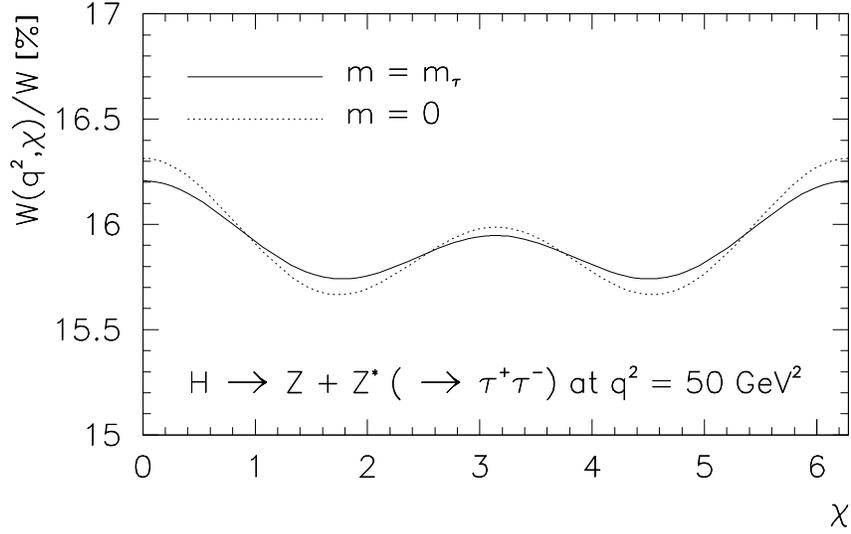, scale=0.7}
\caption{\label{wchi}$\chi$ dependence of the normalized decay distribution
$\widetilde W^{Z}(q^{2},\chi)$ for $q^{2}=50\GeV^{2}$ in case of $m_{\ell}=0$
(dotted line) and $m_{\ell}=m_{\tau}$ (solid line)}
\end{center}\end{figure}

Finally we turn to the normalized single-angle azimuthal distribution, where
\begin{equation}
\widetilde W^{Z}(\chi)=\frac{1}{2\pi}\left(1
  +\frac{\pi^{2}}{16}{\cal\widetilde F}^{Z}_{5}\cos\chi
  +\frac49{\cal\widetilde F}^{Z}_{7}\cos 2\chi\right).
\end{equation}
In Fig.~\ref{wchi} we show the $\chi$ dependence of
$\widetilde W^{Z}(q^{2},\chi)$ again for $q^{2}=50\GeV^{2}$. The nonflip
contribution to the coefficient of the $\cos 2\chi$ term clearly dominates
the decay distribution since the leading contribution is given by
$f^{Z}_{7}\sim a_{\ell}^{2}v_{q}^{2}\rho_{+-}=a_{\ell}^{2}v_{q}^{2}$.
The dominance of the $\cos 2\chi$ term is clearly evident in Fig.~\ref{wchi}.
Lepton-mass effects are generally small and amount to maximally $\sim 1.2 \,\%$
at $\chi=0$, $\pi/2$, $3/2\pi$, $2\pi$ where the mass dependence mainly results
from the normalization.

\subsection{The polarization of the off-shell gauge boson $Z^{\ast}$}
\begin{figure}
\begin{center}
\epsfig{figure=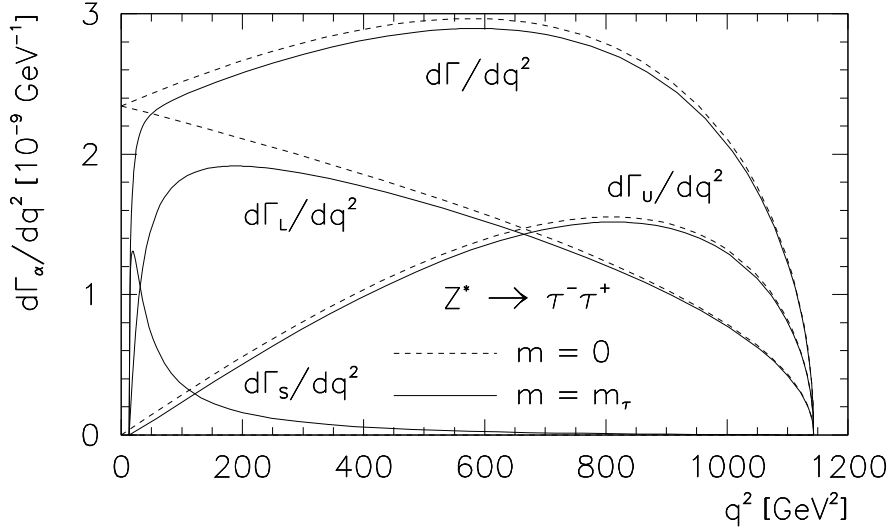, scale=0.7}
\end{center}
\caption{\label{dgamhz}Differential rates $d\Gamma^{Z}_{\alpha}/dq^{2}$
(indices $\alpha=U,L,S$ and vanishing index for $\alpha=U+L$) for the
decay $H\to Z(\to e^{+}e^{-})+Z^{\ast}(\to \ell^{+}\ell^{-})$
with $m_{\ell}=0$ and $m_{\ell}=m_{\tau}$}
\end{figure}

In Sec.~\ref{sec32} we have already considered the single-angle
$\cos\theta_{q}$ distribution which we wrote in the form
$\sim (1+{\cal\widetilde F}^{Z}_{1}P_{2}(\cos\theta_{q}))$. In this subsection 
we want to write the same angular decay distribution in terms of the
transverse, longitudinal and scalar components $\rho_{U}$, $\rho_{L}$ and
$\rho_{S}$ of the double spin-density matrix $\rho_{mm'}$. One obtains
\begin{eqnarray}\label{ULS}
\frac{d\Gamma^{Z}}{dq^{2}d\cos\theta_{q}}
  &=&2p^{2}2q^{2}B_{Z\ell\ell}C^{Z}(q^{2})
  \Bigg\{\left[\frac38(1+\cos^2\theta_{q})\,\rho_{U}
  +\frac34\sin^2\theta_{q}\,\rho_{L}\right]C_{ew}^{(q)}\nn&&
  +\frac{2m_{\ell}^{2}}{q^{2}}
  \left[\left(\frac34\sin^2\theta_{q}\,\rho_{U}
  +\frac32\cos^2\theta_{q}\,\rho_{L}\right)v_{\ell}^{2}
  +\frac32F_{S}^{2}(q^{2})\rho_{S}a_{\ell}^{2}\right]\Bigg\}\nn 
  &=&2p^{2}2q^{2}B_{Z\ell\ell}C^{Z}(q^{2})\Bigg\{
  \left[\frac38(1+\cos^2\theta_{q})\rho_{U}+
  \frac34\sin^2\theta_{q}\,\rho_{L}\right]
  \Big(C_{ew}^{(q)}-4\eps v_\ell^{2}\Big)\nn&&
  +3\eps\Big(\rho_{U+L}v_{\ell}^{2}+F_{S}^{2}(q^{2})\rho_{S}a_{\ell}^{2}
  \Big)\Bigg\}.
\end{eqnarray}
Integrating the differential rate~(\ref{ULS}) with respect to
$\cos\theta_{q}$, one obtains
\begin{equation}\label{dgamzULS}
\frac{d\Gamma^{Z}}{dq^{2}}=2p^{2}2q^{2}B_{Z\ell\ell}C^{Z}(q^{2})
\Bigg\{\left(C_{ew}^{(q)}
  +\frac{2m_{\ell}^{2}}{q^{2}}v_{\ell}^{2}\right)(\rho_{U}+\rho_{L})
  +\frac{6m_{\ell}^{2}}{q^{2}}F_{S}^{2}(q^{2})a_{\ell}^{2}\rho_{S}\Bigg\}.
\end{equation}
Equation~(\ref{dgamzULS}) can be seen to be the equivalent of the $i=0$ 
piece of Eq.~(\ref{partialz}). Accordingly we define partial rates by writing
\begin{eqnarray}
\frac{d\Gamma^{Z}_{U,L}}{dq^{2}}&=&2p^{2}2q^{2}B_{Z\ell\ell}C^{Z}(q^{2})
  \left(C_{ew}^{(q)}+\frac{2m_{\ell}^{2}}{q^{2}}v_{\ell}^{2}\right)
  \rho_{U,L}\,,\nn
\frac{d\Gamma^{Z}_{S}}{dq^{2}}&=&2p^{2}2q^{2}B_{Z\ell\ell}C^{Z}(q^{2})
\frac{6m_{\ell}^{2}}{q^{2}}F_{S}^{2}(q^{2})a_{\ell}^{2}\rho_{S}.
\end{eqnarray}
In Fig.~\ref{dgamhz} we display the $q^{2}$ dependence of the three partial
rates $d\Gamma^{Z}_{\alpha}/dq^{2}$ ($\alpha=U,L,S$) for the two cases
$m_{\ell}=0$ and $m_{\ell}=m_{\tau}$. Lepton mass effects are largest for
$q^{2}$ values in the vicinity of the threshold. There is a substantial
tauonic scalar rate for low $q^{2}$ values which partially compensates for the
loss of longitudinal rate in the low $q^{2}$ region. In the massless case,
where there is no scalar partial rate, the longitudinal rate dominates the
transverse rate up to $\sim 700\GeV^{2}$. The transverse rate
$d\Gamma^{Z}_{U}/dq^{2}$ shows a very small lepton-mass dependence. The 
transverse rates vanish at threshold due to their respective threshold
factors. The $q^{2}=0$ mass-zero rate can be seen to be in numerical agreement
with the corresponding $q^{2}=0$ value of Ref.~\cite{Keung:1984hn} listed in
Eq.~(\ref{keungnum}).

The angular decay distributions are given by the normalized partial rates
$d\Gamma_{U,L,S}/dq^{2}$ divided by the total rate  $d\Gamma_{U+L+S}/dq^{2}$.
For brevity we denote these normalized rates by $\widetilde U$, $\widetilde L$
and $\widetilde S$. Again we take our reference value $q^{2}=50\GeV^{2}$. The
transverse--longitudinal--scalar helicity fractions are given by
\begin{equation}\label{helfracZ}
\widetilde U\,:\,\widetilde L\,:\,\widetilde S \, =\,0.06\,:\,0.94\,:\,0
 \qquad \qquad \widetilde U\,:\,\widetilde L\,:\,\widetilde S \, 
=\, 0.04\,:\,0.65\,:\,0.31.
\end{equation}
The left and right three values refer to the modes
$H\to Z(\to e^{+}e^{-})+Z^{\ast}(\to\mu^{+}\mu^{-})$ and 
$H\to Z(\to e^{+}e^{-})+Z^{\ast}(\to\tau^{+}\tau^{-})$, respectively.
In the $\tau$ mode one observes a substantial loss in the longitudinal rate
which is compensated for by the appearance of the scalar rate. This has
consequences for the $\cos\theta_{q}$ distribution as shown in Fig.~\ref{wthq}.
 
In the upper part of Table~\ref{tgamhz} we list the total rate and the mean
values of the transverse, longitudinal and scalar partial rates for the
mass-zero modes and the $\tau$ mode where the mean is taken with regard to
$q^{2}$. Lepton-mass effects reduce the total rate by $3.97\,\%$. The rate
reduction is largest for the longitudinal rate where the rate reduction
amounts to $6.99\,\%$. This is partially made up for by the appearance of 
the scalar rate which amounts to $4.14\,\%$. The average transverse rate is 
practically unaffected by lepton mass effects. 

\subsection{Off-shell -- off-shell decays
  $H\to Z^{\ast}(\to\ell^{+}\ell^{-})+Z^{\ast}(\to\tau^{+}\tau^{-})$}
To conclude the section about Higgs boson decays into a pair of $Z$ bosons, 
we briefly consider the case where both $Z$ bosons are off-shell. The
double-differential decay rate for
$H\to Z^{\ast}(\to\ell^{+}_{p}\ell^{-}_{p})\,
Z^{\ast}(\to \ell^{+}_{q}\ell^{-}_{q})$ reads ($\ell_{p}\neq \ell_{q}$)
\begin{eqnarray}\label{offshellZs1}
\lefteqn{\frac{d\Gamma^{Z}}{dp^{2}dq^2}(p^{2},q^{2})
  \ =\ \frac{g^6}{8\cdot192\cdot192\pi^5}\frac{1}{\cos^{6}\theta_{W}}
  \frac{m_{Z}^{2}}{m_H^2}|\vec p_{V}(p^{2},q^{2})|
  v_{p}v_{q}}\nonumber\\&&\strut\times
  \frac1{(p^2-m_Z^2)^2+m_Z^2\Gamma_Z^2}\,\,
  \frac1{(q^2-m_Z^2)^2+m_Z^2\Gamma_Z^2}\, \frac{1}{16}\nonumber\\&&\strut\times
  \Bigg\{L_{1}^{Z}(p^{2})P_{1}^{\mu\nu}(p)
  +3)F_{S}^{2}(p^{2})L_{0}^{Z}(p^{2})
  P_{0}^{\mu\nu}(p)\Bigg\}\nonumber\\&&\strut\times
  \Bigg\{L_{1}^{Z}(q^{2})P_{1}^{\mu\nu}(q)
  +3F_{S}^{2}(q^{2})L_{0}^{Z}(q^{2})
  P_{0}^{\mu\nu}(q)\Bigg\},
\end{eqnarray}
where we have written the result in terms of the spin-1 and spin-0 
projections of the neutral current lepton tensor listed in Appendix~B.
The spin-1 and spin-0 propagators $P_{1}^{\mu\nu}$ and $P_{0}^{\mu\nu}$
are defined in Eq.~(\ref{split}). The velocity-type parameters $v_{p}$ and
$v_{q}$ are defined by
$v_{p}=2|\vec p_{\ell p}|/\sqrt{p^{2}}=\sqrt{1-4m_{\ell p}^{2}/p^{2}}$ and
$v_{q}=2|\vec p_{\ell q}|/\sqrt{q^{2}}=\sqrt{1-4m_{\ell q}^{2}/q^{2}}$.
 
In writing down Eq.~(\ref{offshellZs1}) we have chosen a $p\leftrightarrow q$
symmetric representation. This symmetric form is very useful when one
discusses the identical-particle decay $H\to\tau^{+}\tau^{-}\tau^{+}\tau^{-}$
where two of the four contributing diagrams have the factorizing form of
Eq.~(\ref{offshellZs1}). To achieve the $p\leftrightarrow q$ symmetry one has
to add the scalar pieces to the $p$-side propagators in Eq.~(\ref{angdis1}),
i.e.\ one replaces $P_{1}^{\alpha \mu}(p)$ by $P_{0\oplus1}^{\alpha\mu}(p)$
etc. The representation~(\ref{offshellZs1}) is then obtained by expanding e.g.\
$\int\,d\Omega_{q}\,P_{0\oplus1}^{\alpha\mu}(q)\,L^{(q)}_{\mu\nu}(q)\,
P_{0\oplus1}^{\nu\beta}(q)$ along $P_{1}^{\alpha\beta}$ and
$P_{0}^{\alpha\beta}$. 

The contractions of the propagator factors can be calculated to be 
\begin{eqnarray}
\label{contract11}
P_{1}^{\mu\nu}(p)P_{1\,\mu\nu}(q)
  &:=&\ \rho_{U+L}(p^2,q^2)\ =\ 2+\frac{(pq)^{2}}{p^{2}q^{2}}
=2 +1+\frac{m_{H}^{2}|\vec p_{Z^*}|^{2}}{p^{2}q^{2}},\\
\label{contract10}
P_{1}^{\mu\nu}(p)P_{0\,\mu\nu}(q)
  &:=&\rho_{S}(p^2,q^2)\ =\ -1+\frac{(pq)^{2}}{p^{2}q^{2}}\ 
  \ =\ \frac{m_{H}^{2}|\vec p_{Z^*}|^{2}}{p^{2}q^{2}},\\
\label{contract00}
P_{0}^{\mu\nu}(p)P_{0\,\mu\nu}(q)
  &:=&\rho_{SS}(p^2,q^2)\ =\ \frac{(pq)^{2}}{p^{2}q^{2}}
=1+\frac{m_{H}^{2}|\vec p_{Z^*}|^{2}}{p^{2}q^{2}}. 
\end{eqnarray}
The transverse and longitudinal pieces in Eq.~(\ref{contract11}) are given by
$\rho_{U}=2$ and $\rho_{L}=(pq)^{2}/p^{2}q^{2}$. The scalar--scalar
contribution $\rho_{SS}=P_{0}^{\mu\nu}(p)P_{0\,\mu\nu}(q)$ in
Eq.~(\ref{offshellZs1}) appears multiplied by the product of helicity-flip
factors $m_{\ell}^{2}/p^{2}\cdot m_{\ell}^{2}/q^{2}$ and can be neglected for
all practical purposes. 

In the present case we take $m_{\ell p}=m_{e}$ on the $p$ side and
$m_{\ell q}=m_{\tau}$ on the $q$ side such that the symmetric
appearance of Eq.~(\ref{offshellZs1}) is lost. In particular, we set $v_{p}=1$
and take the $m_{\ell}\to 0$ limit of the first curly bracket in
Eq.~(\ref{offshellZs1}) replacing it by
$(v_{\ell}^{2}+a_{\ell}^{2})P_{1}^{\mu\nu}(p)$.  

\begin{table}[t]
\begin{center}\begin{tabular}{lllll}\hline\noalign{\vskip 2mm}
&$\Gamma^{Z}$&$\Gamma^{Z}_{U}/\Gamma^{Z}$&$\Gamma^{Z}_{L}/\Gamma^{Z}$
  &$\Gamma^{Z}_{S}/\Gamma^{Z}$
  \\\noalign{\vskip 2mm}\hline\noalign{\vskip 2mm}
\multicolumn5{l}{$H\to Z(\to e^{+}e^{-})
  +Z^{\ast}(\to\ell^{+}\ell^{-})$}\\
($m_{\ell}=m_{\mu}$)
  &$1.008\times 10^{-7}\GeV$&$0.4056$&$0.5940$&$0.0004$
  \\\noalign{\vskip 1mm}
($m_{\ell}=m_{\tau}$)
  &$0.968\times 10^{-7}\GeV$&$0.4062$&$0.5525$&$0.0414$
  \\\noalign{\vskip 1mm}\hline\noalign{\vskip1mm}
\multicolumn5{l}{$H\to Z^{\ast}(\to e^{+}e^{-})
  +Z^{\ast}(\to\ell^{+}\ell^{-})$}\\
($m_{\ell}=m_{\mu}$)
  &$2.449\times 10^{-7}\GeV$&$0.3879$&$0.6119$&$0.0002$
  \\\noalign{\vskip 1mm}
($m_{\ell}=m_{\tau}$)
  &$2.405\times 10^{-7}\GeV$&$0.3881$&$0.5936$&$0.0183$
  \\\noalign{\vskip 1mm}\hline
\end{tabular}\end{center}
\caption{\label{tgamhz}
Total and normalized partial decay rates for the four-body decays
$H\to Z(\to e^{+}e^{-})+Z^{\ast}(\to\ell^{+}\ell^{-})$ (upper part) and
$H\to Z^{\ast}(\to e^{+}e^{-})+Z^{\ast}(\to\ell^{+}\ell^{-})$ (lower part)}
\end{table}

In the lower part of Table~\ref{tgamhz} we present our numerical results for
the off-shell -- off-shell case. We list the total exclusive decay rate 
$\Gamma^{Z}$ and the averages of the partial decay rates
$\Gamma^{Z}_{U}/\Gamma^{Z}$, $\Gamma^{Z}_{L}/\Gamma^{Z}$ and
$\Gamma^{Z}_{S}/\Gamma^{Z}$ for $m_{\ell}=m_{\mu}$ (first line) and
$m_{\ell}=m_{\tau}$ (second line) on the $q$ side ($m_{\ell}=m_{\ell q}$). For
the $p$ side we specify to $m_{\ell p}=m_{e}$. The off-shell -- off-shell
rates can be seen to be approximately twice as big as the on-shell --
off-shell rates. The reason is that in the off-shell -- off-shell case one
picks up contributions from the peaking regions on both the $p$ and $q$ side.
The helicity fractions remain practically unchanged except for the scalar
contribution which is reduced by $\sim 50\,\%$. The reason is that the scalar
contribution comes only from the $q$ side whereas the normalizing rate is
approximately doubled. Our result agrees with the result
$\Gamma^{Z}=1.0256\cdot 10^{-5}\GeV$ of Ref.~\cite{Denner:2011mq} within
$1.5\,\%$.

One can undo the smearing in Eq.~(\ref{offshellZs1}) by the zero-width
substitution
\begin{equation}\label{NWA}
\frac1{(p^{2}-m_{V}^{2})^{2}+m_{V}^{2}\Gamma_{V}^{2}}
  \to\frac{\pi}{m_{V}\Gamma_{V}}\delta(p^{2}-m_{V}^{2})
\end{equation}
with $V=Z$. One then obtains
\begin{eqnarray}\label{dgamhzzn}
\lefteqn{\frac{d\Gamma^{Z}_{q}}{dq^{2}}(q^{2})
  \ =\ \frac{g^4}{4\cdot1536\pi^3}\frac{1}{\cos^{4}\theta_{W}}
  \frac{|\vec p_{V}(p^{2},q^{2})|v_{q}}{m_H^2}}\nonumber\\&&
  \strut\times\frac{B_{Z\ell\ell}}{(q^2-m_Z^2)^2+m_Z^2\Gamma_Z^2}\,m_Z^2
  \left\{L^{Z}_{1}(q^2)\rho_{U+L}(m_Z^2,q^2)
  +3L^{Z}_{0}(q^{2})F_{S}^{2}(q^{2})\rho_{S}(m_Z^2,q^2)\right\}\nn[7pt]
  &=&B_{Z\ell\ell}m_Z^2C^Z(q^2)\left\{L^{Z}_{1}(q^2)\rho_{U+L}(m_Z^2,q^2)
  +3L^{Z}_{0}(q^{2})F_{S}^{2}(q^{2})\rho_{S}(m_Z^2,q^2)\right\},
\end{eqnarray}
where $d\Gamma^{Z}_{q}/dq^{2}$ denotes the differential rate into the
$q$--side leptonic mode. As expected, this result coincides with
Eq.~(\ref{dgamzULS}). The result for the (semi-inclusive) three-body
decay $H\to Z+Z^\ast(\to\ell^+\ell^-)$ can be easily obtained from
Eq.~(\ref{dgamhzzn}) by summing over all channels, i.e.\ by skipping the
branching ratio factor $B_{Z\ell\ell}$. The result~(\ref{dgamhzzn}) without
the factor $B_{Z\ell\ell}$ can be seen to coincide with the result of
Ref.~\cite{Keung:1984hn}.

Finally, also the $q$-side secondary decay process can be considered to be
exclusive. Using the decay rate for the decay $Z\to\ell^+\ell^-$ in
Eq.~(\ref{gamZell}) for both the $p$ and $q$ sides, from
Eq.~(\ref{offshellZs1}) one obtains
\begin{eqnarray}\label{onehalf}
\lefteqn{\frac{d\Gamma^{Z}_{pq}}{dp^{2}dq^2}(p^{2},q^{2})
  \ =\ \frac12\,\frac{g^2m_Z^2}{8\pi\cos^{2}\theta_{W}}
  \frac{|\vec p_{V}(p^{2},q^{2})|v_{p}v_{q}}{m_H^2
  (v_\ell^2+a_\ell^2)^2}}\nonumber\\&&\strut\times
  \frac{B_{Z\ell\ell p}\,m_Z\Gamma_Z}{\pi
  \left((p^2-m_Z^2)^2+m_Z^2\Gamma_Z^2\right)}\,
  \frac{B_{Z\ell\ell q}\,m_Z\Gamma_Z}{\pi
  \left((q^2-m_Z^2)^2+m_Z^2\Gamma_Z^2\right)}\,
  \frac{p^{2}q^{2}}{m_{Z}^{4}}\nonumber\\&&\strut\times
  \Bigg\{v_{\ell}^{2}(1+2\frac{m_{\ell p}^{2}}{p^{2}})P_{1}^{\mu\nu}(p)
  +a_{\ell}^{2}v_{p}^{2}P_{1}^{\mu\nu}(p)
  +3a_{\ell}^{2}\cdot 2\frac{m_{\ell p}^{2}}{p^{2}}F_{S}^{2}(p^{2})
  P_{0}^{\mu\nu}(p)\Bigg\}\nonumber\\&&\strut\times
  \Bigg\{v_{\ell}^{2}(1+2\frac{m_{\ell q}^{2}}{q^{2}})P_{1\,\mu\nu}(q)
  +a_{\ell}^{2}v_{q}^{2}P_{1\,\mu\nu}(q)
  +3a_{\ell}^{2}\cdot 2\frac{m_{\ell q}^{2}}{q^{2}}F_{S}^{2}(q^{2})
  P_{0\,\mu\nu}(q)\Bigg\}.
\end{eqnarray}
$\Gamma^{W}_{pq}$ denotes the rate into the exclusive leptonic modes on the
$p$ and $q$ sides, respectively. In the product
$B_{Z\ell\ell p}B_{Z\ell\ell q}$ there are terms diagonal and nondiagonal in
flavour. The nondiagonal terms appear in pairs referring to the same exclusive
channel. Thus one has to divide by a factor of two in Eq.~(\ref{onehalf}) as
concerns the nondiagonal terms. The diagonal terms appear only once in the
product. In the approximation that the interference contributions of the
diagonal terms can be neglected, the factor $1/2$ correctly counts the number
of diagonal terms (see the discussion in Ref.~\cite{Groote:2015}). From the
inclusive point of view the factor $1/2$ in Eq.~(\ref{onehalf}) appropriately
accounts for the identical particle factor of $1/2$ in the rate. 

It is noteworthy that Eq.~(\ref{onehalf}) cannot be obtained in any gauge
without considering the coupling of the off-shell gauge bosons to fermion
pairs. If one calculates the rate for $H\to Z^{\ast}Z^{\ast}$ in the
spin-1 Lorenz gauge (also called Landau gauge), the result has to be
multiplied by the effective factors $p^2/m_Z^2$ and $q^2/m_Z^2$ in order to
obtain Eq.~(\ref{onehalf}).

\begin{figure}
\begin{center}
\epsfig{figure=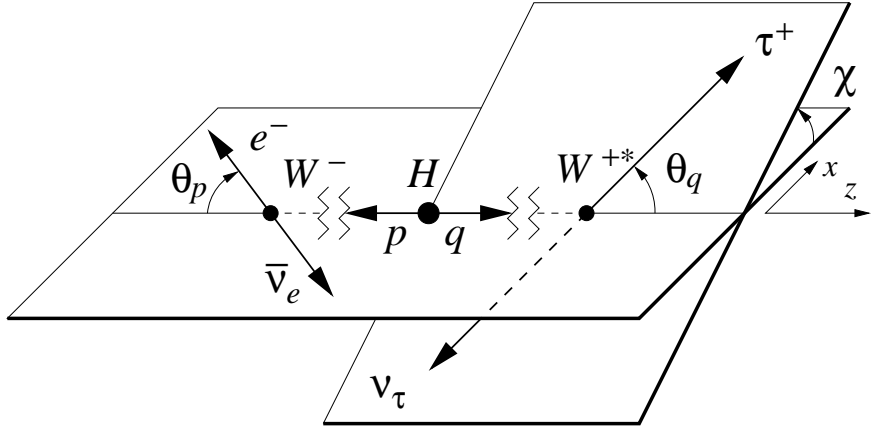, scale=0.7}
\end{center}
\caption{\label{planesww}Definition of the momenta $p$ and $q$, the polar
angles $\theta_{p}$ and $\theta_{q}$, and the azimuthal angle $\chi$ in the
cascade decay $H\to W^{-}(\to \ell^{-}\bar\nu_{\ell})+W^{+\ast}(\to\tau^{+}
\nu_{\tau})$}
\end{figure}

\section{The four-body decay $H\to W^{-}(\to \ell^{-}\bar \nu_{\ell})
+W^{+\ast}(\to \tau^{+}\nu_{\tau})$}
Even though the yield of the ($H \to\ell\nu\ell\nu$) mode from Higgs decay is 
about 40 times larger than the yield of the ($H\to\ell\ell\ell'\ell'$) mode, 
the identification of the $H\to\ell\nu\ell\nu$ mode is much more difficult 
experimentally but can nevertheless be done~\cite{Aad:2015rwa}. In this
section we write down the three-fold angular decay distribution of the cascade
decay
$H\to W^{-}(\to \ell^{-}\bar \nu_{\ell})+W^{+\ast}(\to \tau^{+}\nu_{\tau})$.
In the charged-current decays there are no identical-particle effects such 
that one can also e.g.\ consider the decay
$H\to W^{-}(\to \tau^{-}\bar \nu_{\tau})+W^{+\ast}(\to \tau^{+}\nu_{\tau})$.
As in the neutral-current case the $\tau$ mass can be safely neglected on the
on-shell side since the scale is set by the $W$ mass. At the end of this
section we shall also discuss off-shell -- off-shell decays where the $\tau$
mass can no longer be neglected on the $p$ side. A new feature appearing in
the charged-current decays is the presence of a scalar--longitudinal
interference effect which is parity-conserving but can mimic a
parity-violating contribution. One can anticipate without explicit calculation
that lepton-mass effects are not as important in the charged-current case
since the corresponding helicity-flip contributions are four times weaker than
in the neutral-current case. It is for this reason that we do not discuss the
charged-current case in as much detail as the neutral-current case.

\subsection{Three-fold angular decay distribution for the\\ four-body decay
$H \to W^{-}(\to \ell^{-}\bar\nu_{\ell})+W^{+\ast}(\to \tau^{+}\nu_{\tau})$}  
Let us first consider the three-fold angular decay distribution of the decay
$H \to W^{-}(\to \ell^{-}\bar \nu_{\ell})+W^{+\ast}(\to \tau^{+}\nu_{\tau})$.
The polar angles $\theta_{p}$ and $\theta_{q}$ are defined in the respective
lepton pair center-of-mass systems while the azimutal angle $\chi$ again
describes the relative orientation of the planes as shown in
Fig.~\ref{planesww}. We use the zero lepton-mass approximation for the
on-shell decay $W^{-}\to \ell^{-}\bar \nu_{\ell}$ but keep the $\tau$ mass
finite for the off-shell decay $W^{+\ast}\to \tau^{+}\nu_{\tau}$. 

Again we split the angular decay distribution into its helicity-nonflip and
helicity-flip part. Accordingly we write
\begin{equation}\label{Wjoint2}
W^{W}(\theta_{p},\theta_{q},\chi)=W^{W}_{nf}(\theta_{p},\theta_{q},\chi)
  +W^{W}_{hf}(\theta_{p},\theta_{q},\chi).
\end{equation}
The nonflip decay distribution is given by
\begin{eqnarray}\label{Wjoint3}
\lefteqn{W^{W}_{nf}(\theta_{p},\theta_{q},\chi)
  \ =\ 2p^{2}2q^{2}v_{q}\times\strut}\nn&&
\Big\{(\rho_{++}+\rho_{--})
  \Big[(1+\cos^{2}\theta_{p})(1+\cos^{2}\theta_{q})
  -4\cos\theta_{p}\cos\theta_{q}\Big]
  +4\rho_{00}\sin^{2}\theta_{p}\sin^{2}\theta_{q}\nn&&\strut
  -2(\rho_{++}-\rho_{--})(\cos\theta_{p}-\cos\theta_{q})
  (1-\cos\theta_{p}\cos\theta_{q})\nn&&\strut
  -4(\rho_{+0}+\rho_{-0})\sin\theta_{p}\sin\theta_{q} 
  (1-\cos\theta_{p}\cos\theta_{q})\cos\chi
  +2\rho_{+-}\sin^{2}\theta_{p}\sin^{2}\theta_{q}\cos2\chi\nn&&\strut
  +4(\rho_{+0}-\rho_{-0})\sin\theta_{p}\sin\theta_{q}
  (\cos\theta_{p}-\cos\theta_{q})\cos\chi\Big\}
\end{eqnarray}
($p^{2}=m_{W}^{2}$). For the flip contribution one obtains
\begin{eqnarray}\label{Wjoint4}
\lefteqn{W^{W}_{hf}(\theta_{p},\theta_{q},\chi)
  \ =\ \frac{m_{\tau}^{2}}{q^{2}}\,\cdot2p^{2}2q^{2}v_{q}
  \Big\{(\rho_{++}+\rho_{--})(1+\cos^{2}\theta_{p})
  \sin^{2}\theta_{q}}\nn&&\strut
  +4\rho_{00}\sin^{2}\theta_{p}\cos^{2}\theta_{q}
  -2(\rho_{++}-\rho_{--})\cos\theta_{p}\sin^{2}\theta_{q}\nn&&\strut
  -8F_{S}\rho_{0t}\sin^{2}\theta_{p}\cos\theta_{q}
  +4F_{S}^{2}\rho_{tt}\sin^{2}\theta_{p}
  -(\rho_{+0}+\rho_{-0})\sin2\theta_{p}\sin2\theta_{q}\,\cos\chi\nn&&\strut
  +2(\rho_{+0}-\rho_{-0})\sin\theta_{p}\sin2\theta_{q}\,\cos\chi
  +2F_{S}(\rho_{+t}+\rho_{-t})\sin2\theta_{p}\sin\theta_{q}\,\cos\chi
  \nn&&\strut
  -4F_{S}(\rho_{+t}-\rho_{-t})\sin\theta_{p}\sin\theta_{q}\,\cos\chi
  -2\rho_{+-}\sin^{2}\theta_{p}\sin^{2}\theta_{q}\,\cos2\chi\Big\},
\end{eqnarray}
including the extra minus sign for the spin-1 -- spin-0 interference 
contributions linear in $F_{S}$. We have used the velocity-type parameter
$v_{q}=2|\vec p_{\ell q}|/\sqrt{q^{2}}=1-m_{\ell q}^{2}/q^{2}$.

In Eqs.~(\ref{Wjoint3}) and~(\ref{Wjoint4}) we have also included the
contributions from the parity-violating terms proportional to
$(\rho_{++}-\rho_{--})$, $(\rho_{+0}-\rho_{-0})$ and $(\rho_{+t}-\rho_{-t})$.
These coefficient functions are not populated by the parity-conserving SM
$(HVV)$ coupling. In Appendix~A we briefly discuss the contribution of a
parity-violating non-SM coupling proportional to
$\epsilon^{\mu\nu\rho\sigma}p_{\rho}q_{\sigma}$ which would populate the
$(\rho_{++}-\rho_{--})$, $(\rho_{+0}-\rho_{-0})$ and
$(\rho_{+t}-\rho_{-t})$ coefficient
functions~\cite{Modak:2013sb,Bhattacherjee:2015xra,Zagoskin:2015sca}.   

Again we write the result in terms of the
Legendre polynomials $P_{1}(\cos\theta)=\cos\theta$ and
$P_{2}(\cos\theta)=\frac12(3\cos^2\theta-1)$,
\begin{equation}
(2p^22q^2)^{-1}W^W(\theta_p,\theta_q,\chi)=\frac{16v_{q}}9\sum_{i=0}^{10}
{\cal F}^{W}_ih_i(\theta_p,\theta_q,\chi)=\frac{16v_{q}}9\sum_{i=0}^{10}
(f^{W}_i+\eps g^{W}_i)h_i(\theta_p,\theta_q,\chi).
\end{equation}
The coefficient functions $f^{W}_i$ and $g^{W}_i$ can be found in
Table~\ref{tten} where we have now dropped the non-SM contributions
proportional to $(\rho_{++}-\rho_{--})$, $(\rho_{+0}-\rho_{-0})$ and
$(\rho_{+t}-\rho_{-t})$. It is noteworthy that three new angular structures
proportional to $\cos\theta_q$ ($i=8,9$) and $\sin\theta_q$ ($i=10$) are
generated by a helicity-flip contribution. The first of these contributions
($i=8$) give rise to a nonvanishing forward-backward asymmetry in the
$\cos\theta_{q}$ distribution as discussed later on.

\begin{table}[t]
\begin{center}\begin{tabular}{llll}\hline\noalign{\vskip 2mm}
$i$ \hspace*{0.5cm} & $f^{W}_i$ \hspace*{1.5cm} & $g^{W}_i$ \hspace*{2.5cm} & 
    $h_i(\theta_{p},\theta_{q},\chi)$ 
  \\\noalign{\vskip 2mm}\hline\noalign{\vskip 2mm}
0   & $\rho_{U+L}$
    & $\frac12 \rho_{U+L}\,+\tfrac32F_{S}^{2}\rho_{S}$
    & $1$ 
\\ \noalign{\vskip 1mm}
1   & $\frac12 (\rho_{U}-2\rho_{L})$
    & $-\frac12 (\rho_{U}-2\rho_{L})$
& $P_{2}(\cos\theta_{q})$     
\\ \noalign{\vskip 1mm}
2   & $\frac12 (\rho_{U}-2\rho_{L})$
    & $\frac14 (\rho_{U}-2\rho_{L})$  
$-\tfrac32 F_{S}^{2}\rho_{S}$ & $P_{2}(\cos\theta_{p})$ 
\\ \noalign{\vskip 1mm}
3   & $\frac14 (\rho_{U}+4\rho_{L})$
    & $-\frac14 (\rho_{U}+4\rho_{L})$  &   
      $P_{2}(\cos\theta_{p})P_{2}(\cos\theta_{q})$   
\\ \noalign{\vskip 1mm}
4   & $- \frac94\rho_{U}$  & $0$  &  
$\cos\theta_{p}\cos\theta_{q}$
\\ \noalign{\vskip 1mm}
5   & $-\tfrac{9}{4}(\rho_{+0}+\rho_{-0})$    & $0$  &  
      $\sin\theta_{p}\sin\theta_{q}\cos\chi$ 
\\ \noalign{\vskip 1mm}
6   & $\tfrac{9}{16}(\rho_{+0}+\rho_{-0})$    & $-\tfrac{9}{16}
(\rho_{+0}+\rho_{-0})$  &  
      $\sin2\theta_{p}\sin2\theta_{q}\cos\chi$ 
\\ \noalign{\vskip 1mm}
7  & $\tfrac{9}{8}
\rho_{+-}$    & $-\tfrac{9}{8}
\rho_{+-}$  &  
      $\sin^{2}\theta_{p}\sin^{2}\theta_{q}\cos2\chi$ 
\\ \noalign{\vskip 1mm}
8   & $0$  & $- 3F_{S}\,\rho_{0t}$  & $\cos\theta_{q}$  
\\ \noalign{\vskip 1mm}
9   & $0$  & $ 3F_{S}\,\rho_{0t}$  &  
      $P_{2}(\cos\theta_{p})\cos\theta_{q}$   
\\ \noalign{\vskip 1mm}
10  & $0$    & $\tfrac{9}{8}\,F_{S}
(\rho_{+t}+\rho_{-t})$  &  
      $\sin2\theta_{p}\sin\theta_{q}\cos\chi$ 
\\[1.5ex]
\hline
    \end{tabular}
  \end{center}
\caption{\label{tten}Coefficient functions appearing in the three-fold angular
decay distribution of the decay
$H \to W^{-}(\to \ell^{-}\bar\nu_{\ell})+W^{+\ast}(\to \tau^{+}\nu_{\tau})$.}
\end{table}

We define a normalized decay distribution
\begin{equation}\label{tildeWW}
\widetilde W^{W}(\theta_{p},\theta_{q},\chi)
  =\frac{W^{W}(\theta_{p},\theta_{q},\chi)}{\int W^{W}(\theta'_{p},
  \theta'_{q},\chi')d\cos\theta'_{p}\,d\cos\theta'_{q}\,d\chi'}
  =\frac{1}{8\pi}\Big(1+\sum_{i=1}^{9}{\cal\widetilde F}^{W}_{i}
  h_{i}(\theta_{p},\theta_{q},\chi)\Big),
\end{equation}
where ${\cal\widetilde F}^{W}_i={\cal F}^{W}_i/{\cal F}^{W}_0$
(and $\tilde f^{W}_{i}=f^{W}_{i}/{\cal F}^{W}_{0}$,
$\tilde g^{W}_{i}=g^{W}_{i}/{\cal F}^{W}_0$) and where
\begin{equation}
{\cal F}^{W}_0=f^{W}_{0}+\eps g^{W}_{0}
  =\rho_{U+L}+\tfrac12\eps(\rho_{U+L}+3F_{S}^{2}\rho_{S}).
\end{equation}
The differential decay rate distribution is given by
\begin{equation}
\frac{d\Gamma^W}{dq^2\,d\cos\theta_p\,d\cos\theta_q\,d\chi}
  =B_{W\ell\nu}\frac{C^{W}(q^{2})}{8\pi}\times\frac9{16v_{q}}
  W^{W}(q^{2},\theta_{p},\theta_{q},\chi),
\end{equation}
where
\begin{equation}
C^{W}(q^{2})=\frac{g^{4}v_{q}}{1536\pi^{3}}
  \frac{|\vec p_{V}(m_{W}^{2},q^{2})|v_{q}}{m_{H}^{2}}
  \frac1{(q^{2}-m_{W}^{2})^{2}+m_{W}^{2}\Gamma_{W}^{2}}
\end{equation}
(note the additional factor $v_{q}$ in the numerator) and
$B_{W\ell\nu}=\Gamma_{W\ell\nu}/\Gamma_{W}$. The decay rate for
$W\to\ell\nu$ ($m_\ell=0$) is given by
\begin{equation}\label{gamWell}
\Gamma_{W\ell\nu}=\Gamma(W\to\ell\nu)=\frac{g^2}{48\pi}m_{W}.
\end{equation}
Partial differential rates are defined according to
\begin{equation}\label{partialw}
\frac{d\,\Gamma^{W}_{i}}{dq^{2}}
  =2p^{2}\,2q^{2}B_{W\ell\nu}C^{W}(q^{2}){\cal F}^{W}_{i}(q^{2}).
\end{equation}
The average values $\langle{\cal\widetilde F}^{W}_{i}\rangle$ of the
coefficient functions is given by
\begin{equation}\label{averagew}
\langle{\cal\widetilde F}^{W}_{i}\rangle
=\frac{\int dq^{2}2p^{2}2q^{2}B_{W\ell\nu}C^{W}(q^{2}){\cal F}^{W}_{i}(q^{2})}
{\int dq^{2}2p^{2}2q^{2}B_{W\ell\nu}C^{W}(q^{2}){\cal F}^{W}_{0}(q^{2})}
= \frac{\Gamma^{W}_{i}}{\Gamma^{W}},
\end{equation}
where the integration over $q^{2}$ runs from $4m_{\ell}^{2}$ to
$(m_{H}-m_{W})^{2}$.

\begin{table}
\begin{center}\begin{tabular}{lllll}\hline\noalign{\vskip 2mm}
$i$&${\cal\widetilde F}^{W}_i$ ($m_{\ell}=0$)
&${\cal\widetilde F}^{W}_i$ ($m_{\ell}=m_{\tau}$)
&$\langle{\cal\widetilde F}^{W}_i\rangle$ ($m_{\ell}=0$)
&$\langle{\cal\widetilde F}^{W}_i\rangle$ ($m_{\ell}=m_{\tau}$)
  \\\noalign{\vskip 2mm}\hline\noalign{\vskip 2mm}
 $1$&$-0.9549$&$-0.7983$&$-0.3965$&$-0.3817$\\\noalign{\vskip 2mm}
 $2$&$-0.9549$&$-0.9585$&$-0.3965$&$-0.3985$\\\noalign{\vskip 1mm}
 $3$&$+0.9775$&$+0.8172$&$+0.6983$&$+0.6809$\\\noalign{\vskip 1mm}
 $4$&$-0.6759$&$-0.0603$&$-0.9052$&$-0.9006$\\\noalign{\vskip 1mm}
 $5$&$-0.5431$&$-0.4847$&$-1.4306$&$-1.4205$\\\noalign{\vskip 1mm}
 $6$&$+0.1358$&$+0.1135$&$+0.3576$&$+0.3533$\\\noalign{\vskip 1mm}
 $7$&$+0.0169$&$+0.0141$&$+0.2263$&$+0.2244$\\\noalign{\vskip 1mm}
 $8$&$-0.0000$&$-0.1614$&$-0.0000$&$-0.0176$\\\noalign{\vskip 1mm}
 $9$&$+0.0000$&$+0.1614$&$+0.0000$&$+0.0176$\\\noalign{\vskip 1mm}
$10$&$+0.0000$&$+0.0015$&$+0.0000$&$+0.0029$\\\noalign{\vskip 2mm}
\hline
\end{tabular}\end{center}
\caption{\label{normw}Numerical results for the normalized coefficient
functions ${\cal\widetilde F}^{W}_{i}(q^{2})$ at $q^{2}=50\GeV^{2}$ and the
average of ${\cal\widetilde F}^{W}_{i}(q^{2})$ over
$q^{2}\in[m_{\ell}^{2},(m_{H}-m_{W})^2]$}
\end{table}

In Table~\ref{normw} we present numerical results for the normalized
coefficient functions ${\cal\widetilde F}^{W}_{i}(q^{2})$ and their averages.
In  columns~2 and~3 we list the values of ${\cal\widetilde F}^{W}_{i}(q^{2})$
for $q^{2}=50\GeV^{2}$ with zero and nonzero lepton masses. Lepton-mass
effects amount to $-16\%$ for the functions
${\cal\widetilde F}^{W}_{1,3,6,7}$, $-11.8\%$ for the functions
${\cal\widetilde F}^{W}_{4,5}$, and only $+0.4\%$ for the function
${\cal\widetilde F}^{W}_{2}$. The normalized coefficient functions
${\cal\widetilde F}^{W}_{8,9,10}$ are zero for zero lepton masses. For
$m_{\ell}=m_{\tau}$ the coefficient functions ${\cal\widetilde F}^{W}_{8,9}$
become quite large at $16.1\%$. Again, lepton-mass effects would be even
larger for smaller values of $q^{2}$. Compared to the $H\to ZZ^{\ast}$ case,
lepton mass effects are smaller by approximately a factor of $1/2$.

In columns~4 and~5 of Table~\ref{normw} we also present average values
$\langle{\cal\widetilde F}^{W}_{i}\rangle$ of the coefficient functions again
for zero and nonzero lepton masses where the average is taken with regard to
$q^{2}$. On average lepton mass effects can be seen to be quite small.

\subsection{Forward-backward asymmetry of the $\tau^{+}$ lepton}  
Let us take a closer look at the $\cos\theta_{q}$ distribution determined
by the coefficient functions ${\cal\widetilde F}^{W}_{1}$ and
${\cal\widetilde F}^{W}_{8}$. The normalized $\cos\theta_{q}$ distribution is
given by (see Table~\ref{tten})
\begin{equation}\label{cosdistr}
\widetilde W^{W}(q^{2},\theta_{q})
  =\frac12\left(1+{\cal\widetilde F}^{W}_{1}P_{2}(\cos\theta_{q})
  +{\cal\widetilde F}^{W}_{8}\cos\theta_{q}\right).
\end{equation} 
Contrary to the $H\to ZZ^{\ast}$ case one now has a contribution linear in
$\cos\theta_{q}$ which implies a nonvanishing forward-backward asymmetry. 
It is interesting to note that the source of this parity-odd term in
Eq.~(\ref{fba}) results from a parity-conserving interaction. Consider the
$J^P$ content of the currents coupling to the $W^\ast$: $V^\mu(1^-,0^+)$
and $A^\mu(1^+,0^-)$. The scalar--longitudinal interference contribution
leading to a nonvanishing forward-backward asymmetry can be seen to result
from the parity-conserving interference of the products of currents
$V(0^+)V(1^-)$, $A(0^-)A(1^+)$. We mention that a parity-violating
$(HW^{+}W^{-})$ coupling as discussed in Appendix~A would also give rise to a
nonvanishing forward-backward asymmetry proportional to
$(\rho_{++}-\rho_{--})$ (see Eq.~(\ref{Wjoint3})).

\begin{figure}[t]\begin{center}
\epsfig{figure=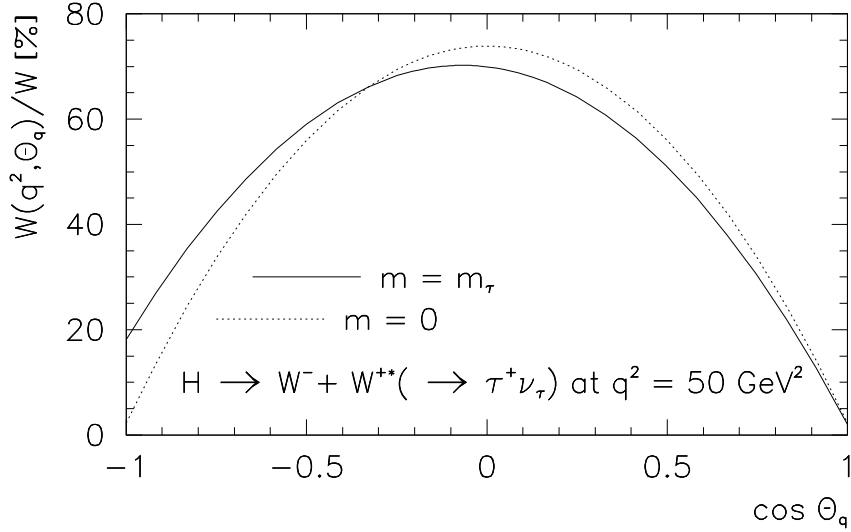, scale=0.7}
\caption{\label{wthqw}$\cos\theta_{q}$ dependence of the normalized decay
distribution $\widetilde W^{W}(q^{2},\theta_q)$ for $q^{2}=50\GeV^{2}$ 
for $m_{\ell}=0$ (dotted line) and $m_{\ell}=m_{\tau}$ (solid line)}
\end{center}\end{figure}

In Fig.~\ref{wthqw} we display the normalized $\cos\theta_{q}$ distribution
for a fixed value of $q^{2}=50\GeV^{2}$. Since $\rho_{U}\ll\rho_{L}$ for
$q^{2}=50$ GeV$^{2}$, the governing feature of the distribution is described
by a downward open parabola. The convexity parameter is proportional to
$(1-\eps)$, leading to a smaller convexity for the $m_{\ell}=m_{\tau}$
distribution as can be seen in Fig.~\ref{wthqw}. There is a pronounced
forward-backward asymmetry in the $\tau$ mode. According to
Eq.~(\ref{cosdistr}) the forward-backward asymmetry of the $\cos\theta_{q}$
distribution is given by
\begin{equation}\label{fba}
A_{FB}(q^{2})=\frac{\Gamma_F-\Gamma_B}{\Gamma_F+\Gamma_B}
  =\frac12{\cal\widetilde F}^{W}_{8}(q^{2})
  =\frac{m_{\ell}^{2}}{q^{2}}\times
  \frac{-3F_{S}(q^{2})\rho_{0t}(q^{2})}{2{\cal F}^{W}_{0}(q^{2})}.
\end{equation}

Since $\rho_{0t}$ and $F_{S}$ are positive, the forward-backward asymmetry 
$A_{FB}$ is negative as also shows up in Fig.~\ref{wthqw}. In fact, one 
calculates 
\begin{equation}
A_{FB}(q^{2}=50\GeV^{2})=-0.081.
\end{equation}
For smaller $q^{2}$ values the forward-backward asymmetry becomes even more
pronounced. In the last column of Table~\ref{tgamhw} we list the average value
of $A_{FB}$ which is given by
$\langle A_{FB}\rangle=\frac12\langle{\cal\widetilde F}^{W}_{8}\rangle$.
On average, the forward-backward asymmetry is much smaller than for
$q^{2}=50\GeV^{2}$.

\subsection{The polarization of the  off-shell $W^{+\ast}$}
\begin{figure}
\begin{center}
\epsfig{figure=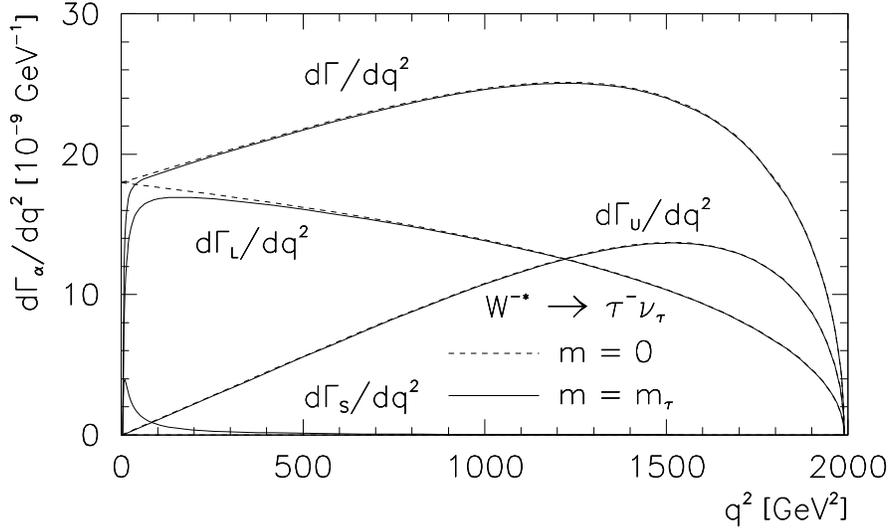, scale=0.7}
\end{center}
\caption{\label{dgamhw}Differential rates $d\Gamma^{W}_{\alpha}/dq^{2}$
(indices $\alpha=U,L,S$ and vanishing index for $\alpha=U+L$) for the
decay $H\to W^{-}(\to e^{-}\bar\nu_e)+W^{+\ast}(\to\ell^{+}\nu_\ell)$
with $m_{\ell}=0$ and $m_{\ell}=m_{\tau}$}
\end{figure}

Let us rewrite the single-angle $\cos\theta_{q}$ distribution in
Eq.~(\ref{cosdistr}) in terms of the transverse, longitudinal, scalar and the
scalar--longitudinal interference contributions. One has 
\begin{eqnarray}\label{ULSW}
\lefteqn{\frac{d\Gamma^{W}}{dq^{2}d\cos\theta_{q}}\ =\ 2p^{2}2q^{2}B_{W\ell\nu}
  C^{W}(q^{2})\Bigg\{\frac38(1+\cos^2\theta_{q})\rho_{U}
  +\frac34\sin^2\theta_{q}\rho_{L}}\nn&&
  +\frac{m_{\ell}^{2}}{2q^{2}}\left[\frac34\sin^2\theta_{q}\rho_{U}
  +\frac32\cos^2\theta_{q}\rho_{L}-3F_{S}(q^{2})\cos\theta_{q}\rho_{t0}
  +\frac32F_{S}^{2}(q^{2})\rho_S\right]\Bigg\}.
\end{eqnarray}
Integrating the differential rate~(\ref{ULS}) with respect to
$\cos\theta_{q}$, one obtains
\begin{equation}\label{dgamwULS}
\frac{d\Gamma^{W}}{dq^{2}}=2p^{2}\,2q^{2}\,B_{W\ell\nu}C^{W}(q^{2})
  \left\{(\rho_{U}+\rho_{L})+\frac{m_{\ell}^{2}}{2q^{2}}\left[(\rho_{U}
  +\rho_{L})+3F_{S}^{2}(q^{2})\rho_{S}\right]\right\}.
\end{equation}
Partial decay rates are accordingly defined by
\begin{eqnarray}
\frac{d\Gamma^{W}_{U,L}}{dq^{2}}&=&2p^{2}\,2q^{2}\,B_{W\ell\nu}C^{W}(q^{2})
  \left(1+\frac{m_{\ell}^{2}}{2q^{2}}\right)\rho_{U,L}\,,\nn
\frac{d\Gamma^{W}_{S}}{dq^{2}}&=&2p^{2}\,2q^{2}\,B_{W\ell\nu}C^{W}(q^{2})
  \frac{3m_{\ell}^{2}}{2q^{2}}\,F_{S}^{2}(q^{2})\rho_{S}.
\end{eqnarray}
In Fig.~\ref{dgamhw} we display the $q^{2}$ dependence of the partial decay
rates. Lepton-mass effects show up mainly for lower $q^{2}$ values between
threshold and approximately $150\GeV^{2}$. For example, at $q^{2}=50\GeV^{2}$
the helicity fractions of the $W^{+\ast}$ change according to 
\begin{equation}\label{helfracW}
\widetilde U\,:\,\widetilde L\,:\,\widetilde S
  \,=\,0.030\,:\,0.970\,:\,0\qquad\to\qquad 
\widetilde U\,:\,\widetilde L\,:\,\widetilde S
  \,=\,0.028\,:\,0.893\,:\,0.079
\end{equation}
when going from the $e,\mu$ modes to the $\tau$ mode. The picture is similar
to the $H\to ZZ^{\ast}$ case. However, lepton-mass effects are less pronounced
in the $H \to WW^{\ast}$ case.

In the upper part of Table~\ref{tgamhw} we list the total rate and the mean
values of the transverse, longitudinal and scalar partial rates for the
mass-zero modes and the $\tau$ mode where the mean is taken with regard to
$q^{2}$. In this case, lepton-mass effects reduce the total rate by $0.76\,\%$.
The rate reduction is largest for the longitudinal rate where the rate
reduction amounts to $2.1\,\%$. On average, the rate reduction for
$\langle\,\Gamma_{L}\rangle = \Gamma_{L}/\Gamma$ is still a considerable
$1.3\,\%$ while the average of the transverse rate is practically unchanged.
As in the $W\to ZZ^{\ast}$ case, the loss of longitudinal rate is mainly
compensated for by the appearance of the scalar rate. Our result agrees
with the result $\Gamma^{W}=2.4135\cdot 10^{-7}\GeV$ of
Ref.~\cite{Denner:2011mq} within $1.6\,\%$.

\subsection{Off-shell -- off-shell decays 
$H \to W^{-\ast}(\to \ell^{-}\bar\nu_{\ell})
+W^{+\ast}(\to \tau^{+}\nu_{\tau})$}
Again, we conclude the section by considering the case where both $W$ bosons
are off-shell. Using again the narrow-width approximation~(\ref{NWA}) for
$V=W$ also on the $p$ side, the exclusive off-shell -- off-shell rate is
obtained from the double integral
\begin{equation}\label{offoffrate}
\Gamma^{W}_{pq}=\int_{m^{2}_{\ell p}}^{m^{2}_{H}}dp^{2}
\frac{B_{W\ell\nu p}\,m_{W}\Gamma_{W}}{\pi((p^2-m_W^2)^2+m_W^2\Gamma_W^2)}
\int_{m^{2}_{\ell q}}^{(m_{H}-\sqrt{p^{2}})^{2}}dq^{2}
\frac{B_{W\ell\nu q}\,m_{W}\Gamma_{W}}{\pi((q^2-m_W^2)^2+m_W^2\Gamma_W^2)}
\Gamma^{W}_{0},
\end{equation}
where $\Gamma^{W}_{pq}$ denotes the rate into the exclusive leptonic modes
on the $p$ and $q$ sides. The total inclusive mode is obtained by summing over
all exclusive modes including the quark--antiquark modes, i.e.\ by setting
$B_{W\ell\nu p}=B_{W\ell\nu q}=1$. The rate function $\Gamma^{W}_{0}$ in
Eq.~(\ref{offoffrate}) reads
\begin{eqnarray}\label{wrate}
\Gamma^{W}_{0}&=&\frac{g^{2}}{8\pi}|\vec {p}_{V}(p^{2},q^{2})|
\frac{1}{m_{W}^{2}m^{2}_{H}}\,\frac{v_{p}v_{q}}{64}
\Bigg\{L_{1}^{W}(p^{2})P_{1}^{\mu\nu}(p)+3F_{S}^{2}(p^{2})L_{0}^{W}(p^{2})
  P_{0}^{\mu\nu}(p)\Bigg\}\nonumber\\&&\strut\qquad\qquad\qquad\qquad\times
  \Bigg\{L_{1}^{W}(q^{2})P_{1}^{\mu\nu}(q)+3F_{S}^{2}(q^{2})L_{0}^{W}(q^{2})
  P_{0}^{\mu\nu}(q)\Bigg\},
\end{eqnarray}
where $v_{p}=1-m_{\ell p}^{2}/p^{2}$ and $v_{q}=1-m_{\ell q}^{2}/q^{2}$.
$L^{W}_{1}$ and $L^{W}_{0}$ are the spin-1 and spin-0 projections of the
charged current lepton tensors listed in Appendix~C. When the lepton masses
are taken to be zero, the rate function~(\ref{wrate}) simplifies to
\begin{equation}
\Gamma^{W}_{0}(m_{\ell}=0)=\frac{g^{2}}{8\pi}|\vec {p}_{W}|
\frac{p^{2}q^{2}}{m_{W}^{2}m^{2}_{H}}\,\rho_{U+L}
\end{equation}
which agrees with the results in Refs.~\cite{Djouadi:2005gi,Grau:1990uu}.

\begin{table}[t]
\begin{center}\begin{tabular}{llllll}\hline\noalign{\vskip 2mm}
&$\Gamma^{W}$&$\Gamma^{W}_{U}/\Gamma^{W}$&$\Gamma^{W}_{L}/\Gamma^{W}$
  &$\Gamma^{W}_{S}/\Gamma^{W}$&$\langle A_{FB}\rangle$
  \\\noalign{\vskip 2mm}\hline\noalign{\vskip 2mm}
\multicolumn5{l}{$H\to W^{-}(\to e^{-}\bar\nu_e)
  +W^{+\ast}(\to\ell^{+}\nu_\ell)$}\\
($m_\ell=m_\mu$)
  &$4.754\times 10^{-6}\GeV$&$0.4023$&$0.5976$&$0.0001$
  &$-5\times 10^{-5}$\\\noalign{\vskip 1mm}
($m_\ell=m_\tau$)
  &$4.718\times 10^{-6}\GeV$&$0.4033$&$0.5894$&$0.0073$&$-0.0059$
  \\\noalign{\vskip 1mm}\hline\noalign{\vskip1mm}
\multicolumn5{l}{$H\to W^{-\ast}(\to e^{-}\bar\nu_e)
  +W^{+\ast}(\to\ell^{+}\nu_\ell)$}\\
($m_\ell=m_\mu$)
  &$1.009\times 10^{-5}\GeV$&$0.3952$&$0.6048$&$3\times 10^{-5}$
  &$-2\times 10^{-5}$\\\noalign{\vskip 1mm}
($m_\ell=m_\tau$)
  &$1.005\times 10^{-5}\GeV$&$0.3956$&$0.6009$&$0.0035$&$-0.0029$
  \\\noalign{\vskip 1mm}\hline
\multicolumn5{l}{$H\to W^{-\ast}(\to \ell^{-}\bar\nu_\ell)
  +W^{+\ast}(\to \ell^{+}\nu_\ell)$}\\
($m_\ell=m_\mu$)
  &$1.009\times 10^{-5}\GeV$&$0.3952$&$0.6048$&$6\times 10^{-5}$
  &$-6\times 10^{-5}$\\\noalign{\vskip 1mm}
($m_\ell=m_\tau$)
  &$1.001\times 10^{-5}\GeV$&$0.3960$&$0.5970$&$0.0070$&$-0.0076$
  \\\noalign{\vskip 1mm}\hline
\end{tabular}\end{center}
\caption{\label{tgamhw}Total and normalized partial decay rates and the
average value of the forward-backward symmetry for the four-body decays
$H\to W^{-}(\to e^{-}\bar\nu_e)+W^{+\ast}(\to\ell^{+}\nu_\ell)$ (first part),
$H\to W^{-\ast}(\to e^{-}\bar\nu_e)+W^{+\ast}(\to\ell^{+}\nu_\ell)$ (second
part), and
$H\to W^{-\ast}(\to \ell^{-}\bar\nu_\ell)+W^{+\ast}(\to\ell^{+}\nu_\ell)$
(third part)}
\end{table}

In Table~\ref{tgamhw} we have listed our numerical results for the off-shell
-- off-shell rates and the averages of the polarization of the gauge bosons.
The off-shell -- off-shell rates approximately amount to two times the
off-shell -- on-shell rates. The reason is again that one picks up
contributions from the peaking regions both on the $p$ side and on the $q$
side. Put in a different language the off-shell -- off-shell
rate~(\ref{offoffrate}) corresponds to the sum of the two off-shell --
on-shell rates $H\to W^{-}W^{+\ast}$ and $H\to W^{-\ast}W^{+}$. The
polarizations listed in Table~\ref{tgamhw} are thus an average of the
respective polarizations on the $p$ side and the $q$ side. This also explains
the fact that the scalar polarization is only one-half of the on-shell --
off-shell value since it is contributed to only by the $q$ side
$H\to W^{-}W^{+\ast}$ channel. This is no longer true for the 
$H\to W^{-\ast}(\to\tau^{-}\bar\nu_\tau)+W^{+\ast}(\to\tau^{+}\nu_\tau)$ mode
where the scalar rate obtains contributions from both the $p$ side and the $q$
side.

The off-shell -- off-shell rates slightly exceed twice the off-shell --
on-shell rates which provides a measure of the quality of taking the
zero-width approximation on the on-shell sides. Compared to the rates for
$H\to Z^{\ast}Z^{\ast}$ listed in Table~\ref{tgamhz}, lepton-mass effects can
be seen to be more than five times smaller in the $H\to W^{-\ast}W^{+\ast}$
case.     

\section{Summary and conclusions}
We have discussed lepton-mass effects in the rate and the angular decay
distributions in the four-body decays
$H\to Z(\to \ell^{+} \ell^{-})+Z^{\ast}(\to \tau^{+}\tau^{-})$ and
$H\to W^{-}(\to \ell^{-}\bar \nu_{\ell})+W^{+\ast}(\to \tau^{+}\nu_{\tau})$
where the gauge bosons $Z$ and $W^{-}$ are on their mass shell. Lepton-mass
effects are larger for the $H \to ZZ^{\ast}$ mode where we find a reduction of
$3.97\,\%$ in the $\tau$ rate relative to the $e,\mu$ rates. In the
$H \to WW^{\ast}$ case the rate reduction of the total rate is smaller.
In this mode we find a rate reduction of $0.76\,\%$ relative to the zero mass
case. Differentially, the rate reduction through lepton-mass effects is 
significantly larger at the lower end of the $q^{2}$ spectrum in both cases.
For both modes we find a significant reduction of the longitudinal rate
through lepton-mass effects in the lower $q^{2}$ region from threshold
to $\sim 200\GeV^{2}$. In this region the transverse--longitudinal composition
of the off-shell gauge bosons is considerably changed. The reduction of the
longitudinal rate in this region is partly compensated for by a significant 
scalar contribution. In the charged-current case one finds a nonvanishing 
forward-backward asymmetry in the $\cos\theta_{q}$ distribution through
lepton-mass induced scalar--longitudinal interference effects. The
forward-backward asymmetry can become quite large in the low-$q^{2}$ region.

We have also discussed the case when both gauge-bosons go off-shell. Double
smearing with the appropriate Breit--Wigner functions increases the overall
rate. Lepton-mass effects become weaker in the double smearing process.

We have employed helicity methods in our analysis which has allowed us to
present our analytical results for angular decay distributions and partial
rates in compact form. In particular, the inclusion of lepton-mass effects in
the helicity formalism is straightforward.

Experimentally it will not be so simple to identify the $\tau$ modes in the
four-lepton decays of the Higgs. In this context we mention that the detection
efficiency for $\tau$ leptons in their hadronic decay channels at the LHC is
being continuously improved 
(see Refs.~\cite{Zinonos:2014rma,Sakurai:2014kha,Jeans:2015vaa}).
Nevertheless, an accurate Monte Carlo event generator for decays involving the
$\tau$ leptons should include lepton-mass effects for which we have supplied
the appropriate matrix elements in this paper. This would e.g.\ be relevant
for modelling $Z\to\tau\tau$ processes as background for the search for the
decay $H\to\tau^{+}\tau^{-}$~\cite{Aad:2015kxa,Bredenstein:2006rh,%
Bredenstein:2006ha}.

\subsection*{Acknowledgments}
This work was supported by the Estonian Research Council under Grant
No.~IUT2-27. J.G.K.\ would like to thank A.~Denner, B.~J\"ager, M.~Rauch,
H.~Spiesberger, B.~Stech and J.~Wang for useful discussions. S.G.\
acknowledges the support by the Mainz Institute of Theoretical Physics (MITP). 

\begin{appendix}

\section{Helicity amplitudes for $H\to VV^{\ast}$}
\setcounter{equation}{0}\def\theequation{A\arabic{equation}}
The helicity amplitudes $H_{\lambda_{V}\lambda_{V^{*}}}$ for the transition 
$H\to VV^{\ast}$ are 
defined by
\begin{equation}
H_{mn}=\bar \eps^{\ast\,\alpha}(\lambda_{V},p)H_{\alpha\alpha'}
\eps^{\ast\,\alpha'}(\lambda_{V^{\ast}},q),\qquad
\end{equation}
where $\bar \eps^{\ast\,\alpha}(\lambda_{V},p)$ and 
$\eps^{\ast\,\alpha'}(\lambda_{V^{\ast}},q)$ are the respective
polarization four-vectors on the $p$ side (on-shell side) and on the $q$ side
(off-shell side). We shall evaluate the helicity amplitudes in the Higgs rest
frame with the $z$ direction defined by the direction of the off-shell
$V^{\ast}$ boson. One therefore has to rotate the polarization four-vectors on
the $p$ side by $180^{\circ}$ which we indicate by the ``bar'' symbol. 

The respective polarization four-vectors in the Higgs rest frame are given by 
\begin{eqnarray}
\mbox {on-shell side}\,&:&\qquad
\bar \eps^{\alpha}(\pm,p)=\frac{1}{\sqrt{2}}(0;\pm1,-i,0) \qquad
\bar \eps^{\alpha}(0,p)=\frac{1}{m_{V}}(|\vec p_{V}|;0,0,-p_{0})\quad
\nn
\mbox {off-shell side}\,&:&\qquad
\eps^{\mu}(\pm,q)=\frac{1}{\sqrt{2}}(0;\mp1,-i,0) \qquad 
\eps^{\mu}(0,q)=\frac{1}{\sqrt{q^{2}}}(|\vec p_{V}|;0,0,q_{0})
\nn
&&\qquad \eps^{\mu}(t,q)=\frac{1}{\sqrt{q^{2}}}
(q_{0};0,0,|\vec p_{V}|) \nonumber
\end{eqnarray}
where $p_{0}=(m^{2}_{H}+p^{2}-q^{2})/(2m_{H})$,  
$q_{0}=(m^{2}_{H}+q^{2}-p^{2})/(2m_{H})$ and 
\begin{equation}
|\vec p_{V}|=\frac{1}{2m_{H}}
  \lambda^{1/2}(m_{H}^{2},p^{2},q^{2})
  =\frac{\sqrt{(pq)^{2}-p^{2}q^{2}}}{m_{H}}
\end{equation}
such that $p^{\mu}=(p_0;0,0,-|\vec p_{V}|)$, $q^{\mu}=(q_0;0,0,|\vec p_{V}|)$.
It is convenient to avail of the covariant representations of the longitudinal
and scalar polarization four-vectors. They read
\begin{eqnarray}
\mbox {on-shell side}\,:
&&\bar \eps^{\mu}(0)=\frac{1}{\sqrt{p^{2}}\sqrt{(pq)^{2}-p^{2}q^{2}}}
\left((pq)p^{\mu}-p^{2}q^{\mu}\right)\qquad
\bar\eps^{\mu}(t)=\frac{p^{\mu}}{\sqrt{p^{2}}}\nn
\mbox {off-shell side}\,:
&&\eps^{\mu}(0)=\frac{1}{\sqrt{q^{2}}\sqrt{(pq)^{2}-p^{2}q^{2}}}
\left((pq)q^{\mu}-q^{2}p^{\mu}\right) \qquad
\eps^{\mu}(t)=\frac{q^{\mu}}{\sqrt{q^{2}}}\qquad
\end{eqnarray}
The helicity amplitudes can then be calculated to be
\begin{equation}
H_{++}=H_{--}=1,\qquad 
H_{00}=H_{tt}=\frac{pq}{\sqrt{p^{2}}\sqrt{q^{2}}},\qquad
H_{0t}=H_{t0}=\frac{\sqrt{(pq)^{2}-p^{2}q^{2}}}{\sqrt{p^{2}}\sqrt{q^{2}}}
  =\frac{m_{H}|\vec p_{V}|}{\sqrt{p^{2}}\sqrt{q^{2}}}.
\end{equation}

The coefficient functions ${\cal F}^{Z,W}_{i}$ are written in terms of
bilinear forms of the helicity amplitudes for which we choose the following
abbreviations
\begin{eqnarray}
\rho_{00}&=&|H_{00}|^{2},\qquad
\rho_{\pm\pm}\ =\ |H_{\pm \pm}|^{2},\qquad
\rho_{tt}\ =\ \real H_{0t}H^{\ast}_{0t},\nn 
\rho_{\pm0}&=&\real H_{\pm\pm}H^{\ast}_{00},\qquad
\rho_{\pm\mp}\ =\ \real H_{\pm\pm}H_{\mp\mp}^{\ast},\nn
\rho_{\pm t}&=&\real H_{\pm\pm}H_{0t}^{\ast},\qquad
\rho_{t \pm}\ =\ \real H_{0t}H_{\pm\pm}^{\ast},\nn
\rho_{0t}&=&\real H_{00}H_{0t}^{\ast},\qquad
\rho_{t0}\ =\ \real H_{t0}H_{00}^{\ast}.
\end{eqnarray}
We sometimes refer to these bilinear forms as the double spin-density matrix
elements of the gauge boson pair since the bilinear forms describe the
entangled polarizations components of the gauge boson pair. The SM values of
the double density matrix $\rho_{mm'}$ are
given by
\begin{eqnarray}\label{dmwstar}
\rho_{++}\ =\ \rho_{--}&=&1,\qquad\qquad
\rho_{\pm\mp}\ =\ 1,\qquad 
\rho_{00}\ =\ \frac{(pq)^{2}}{p^{2}q^{2}}
  \ =\ \Big(1+\frac{m_{H}^2}{q^2m_V^2}|\vec{p}_{V}|^2\Big),\nn
\rho_{0t}\ =\ \rho_{t0}&=&\frac{pq\,\sqrt{(pq)^{2}-p^{2}q^{2}}}{p^{2}q^{2}}
  \ =\ \frac{m_{H}|\vec{p}_{V}|}{2m_V^2q^2}\left(m_{H}^2-m_V^2-q^2\right),\nn 
\rho_{tt}&=&\frac{(pq)^{2}-p^{2}q^{2}}{p^{2}q^{2}}
  \ =\ \frac{m_{H}^2}{q^2m_V^2}|\vec{p}_{V}|^2.
\end{eqnarray}
One notes the following SM relations
\begin{equation}
\rho_{tt}=\rho_{00}-1,\qquad
\rho_{0t}=\sqrt{\rho_{00}\rho_{tt}},\qquad
\rho_{\pm 0}=\rho_{\pm t}=\sqrt{\rho_{00}}.
\end{equation}
At maximal recoil where $q^2$ and/or $p^2$ tend to zero, the dominant double
spin-density matrix elements are $\rho_{00}=\rho_{0t}=\rho_{t0}=\rho_{tt}$.
At minimal recoil $q^2=(m_H-m_V)^2$ where $|\vec p_V|\to 0$, the dominant
contributions are $\rho_{++}=\rho_{--}=\rho_{00}$ while $\rho_{0t}$ and
$\rho_{tt}$ tend to zero.

Some authors prefer to use Cartesian components for the transition matrix
elements~\cite{Bhattacherjee:2015xra} instead of the helicity components used 
by us. The relation between the two representations is given by
\begin{eqnarray}
A_{\parallel}&=&\frac{1}{\sqrt{2}}(H_{++}+H_{--}),\qquad
A_{\perp}\ =\ \frac{1}{\sqrt{2}}(H_{++}-H_{--}),\qquad
A_{0}\ =\ H_{00}.
\end{eqnarray}
We see no particular advantages to write the angular coefficient functions
in terms of their Cartesian components.

Battacherjee {\it et al.} considered two additional non-SM $(HVV)$ coupling
structures~\cite{Bhattacherjee:2015xra}. They write down the effective
coupling structure
\begin{equation}
V^{\mu\nu}=ag^{\mu\nu}+b(q^{\mu}p^{\nu}-(pq)g^{\mu\nu})
  +ic\epsilon^{\mu\nu\rho\sigma}p_\rho q_\sigma.
\end{equation}
The helicity components are then given by
\begin{equation}
H_{00}=a\frac{pq}{\sqrt{p^2}\sqrt{q^2}}-b\sqrt{p^2}\sqrt{q^2},\qquad
H_{\pm\pm}=a\mp c\sqrt{(pq)^2-p^2q^2}.
\end{equation}
There are no contributions of the new coupling structures to $H_{0t}$. It is
clear that one now has a contribution to the difference $(H_{++}-H_{--})$
resulting from the parity-violating term proportional to 
$\epsilon^{\mu\nu\rho\sigma}p_{\rho}q_{\sigma}$.

\section{Helicity representation of the\\ neutral-current lepton tensor}
\setcounter{equation}{0}\def\theequation{B\arabic{equation}}
We calculate the helicity representation of the neutral-current lepton tensor
on the off-shell $q$ side. The corresponding expressions for the on-shell
$p$ side can be obtained by setting the lepton mass to zero and replacing
$q \to p$. We work in the center-of-mass system of the lepton pair with the 
$z$ direction defined by $\ell^{+}$ which we refer to as the helicity system.
The kinematics in the helicity system is given by
\begin{eqnarray}\label{qside}
q^{\alpha}=\sqrt{q^{2}}\,(1;0,0,0)&&
\ell^{\pm\alpha}=\tfrac12\sqrt{q^{2}}(1;0,0,\pm v_{q})\nn
\eps^{\mu}(0)=(0;0,0,1)&&
\eps^{\mu}(\pm)=\frac{1}{\sqrt{2}}(0;\mp1,-i,0)\qquad
\eps^{\mu}(t)=(1;0,0,0)
\end{eqnarray}
where $v_{q}^2=1-4m_q^2/q^2=1-4\eps$. The covariant forms of the lepton
tensors read
\begin{eqnarray}
L_{\mu\nu}^{VV}&=&\Tr\left(\gamma_{\mu}(\slell^{+}-m_q)\gamma_{\nu}
  (\slell^{-}+m_q)\right)
  \ =\ 4\left(\ell^{+}_{\mu}\ell^{-}_{\nu}+\ell^{+}_{\nu}\ell^{-}_{\mu}
  -\tfrac12 q^{2}g_{\mu\nu}\right),\nn
L_{\mu\nu}^{AA}&=&\Tr\left(\gamma_{\mu}\gamma_{5}(\slell^{+}-m_q)
  \gamma_{\nu}\gamma_{5}(\slell^{-}+m_q)\right)
  \ =\ 4\left(\ell^{+}_{\mu}\ell^{-}_{\nu}+\ell^{+}_{\nu}\ell^{-}_{\mu}
  -\tfrac12(q^{2}-4m_q^{2})g_{\mu\nu}\right),\nn
L_{\mu\nu}^{VA}&=&L_{\mu\nu}^{AV}=\Tr\left(\gamma_{\mu}(\slell^{+}-m_q)
  \gamma_{\nu}\gamma_{5}(\slell^{-}+m_q)\right)
  \ =\ -4i\epsilon_{\mu\nu\rho\sigma}q^\rho\ell^{+\sigma}.
\end{eqnarray}
The total neutral current lepton tensor is composed according to 
\begin{equation}
L_{\mu\nu}=v^{2}_{\ell}L_{\mu\nu}^{VV}-2v_{\ell}a_{\ell}L_{\mu\nu}^{VA}
+a^{2}_{\ell}L_{\mu\nu}^{AA},
\end{equation}
where the neutral current is defined by 
$J^{\mu}=\bar\psi\gamma^{\mu}(v_{\ell}-a_{\ell}\gamma_{5})\psi$ with
\begin{equation}
v_{\ell}=-1+4\sin^{2}\theta_{W},\qquad 
a_{\ell}=-1\qquad\mbox{for }\ell=e,\mu,\tau.
\end{equation}

In order to calculate the helicity representation of the lepton tensors
in the helicity system one needs to evaluate
\begin{equation}\label{helrep}
\widehat L_{mm'}^{(p)}
= L^{(p)}_{\mu\nu}\eps^{\mu}(m)\eps^{\ast\,\nu}(m'),\qquad
\widehat L_{nn'}^{(q)}
= L^{(q)}_{\mu\nu}\eps^{\mu}(n)\eps^{\ast\,\nu}(n').\nonumber
\end{equation}
All objects referring to the helicity system are denoted by a hat symbol. The
contractions are done in the helicity system using the
representations~(\ref{qside}). Using the explicit forms~(\ref{qside}) one
calculates
\begin{eqnarray}
\widehat L^{VV(q)}_{\pm\pm}&=&2q^2,\quad 
\widehat L^{AA(q)}_{\pm\pm}\ =\ 2q^2(1-4\eps),\quad  
\widehat L_{\pm\pm}^{VA(q)}=\widehat L_{\pm\pm}^{AV(q)}
  \ =\ \mp 2q^{2}(1-4\eps)^{1/2},\nn
\widehat L^{VV(q)}_{00}\ &=&\ 8q^2\eps,\quad
\widehat L^{AA(q)}_{00}=0,\quad
\widehat L_{tt}^{VV(q)}\ =\ 0, \nn
\widehat L_{tt}^{AA(q)}\ &=&\ 8q^2\eps,\quad
\widehat L_{t0}^{VV,AA(q)}\ =\ \widehat L_{0t}^{VV,AA(q)}=0.
\end{eqnarray}
The ratio of helicity-flip and helicity-nonflip contributions can be seen
to be given by 
$\hat L^{VV}_{00}/\hat L^{VV}_{\pm\pm}=4\eps$ and
$\hat L^{AA}_{tt}/\hat L^{AA}_{\pm\pm}=4\eps/(1-4\eps)$.

The components of the off-shell lepton tensor $L_{\lambda_{Z}\,\lambda'_{Z}}$
in the Higgs decay system ($z$ axis along the $Z^{\ast}$ direction) needed in
the evaluation of Eq.~(\ref{angdis2}) can be obtained by rotating the
components of the lepton tensor
$\widehat L_{\hat\lambda_{Z}\,\hat\lambda'_{Z}}$ defined in the helicity 
system according to (double indices are summed)
\begin{equation}
L_{\lambda_{Z}\,\lambda'_{Z}}(q^{2},\cos\theta_{q},\chi)=\sum_{J=0,1}\,
d^{J}_{\lambda_{Z}\,\hat \lambda_{Z}}(\theta_{q})\,
d^{J}_{\lambda'_{Z}\,\hat \lambda'_{Z}}(\theta_{q})
e^{i(\lambda_{Z}-\lambda'_{Z})\chi}
\,\widehat L_{\hat \lambda_{Z}\,\hat \lambda'_{Z}}(q^{2}),
\end{equation}
where Wigner's $d$ functions are given by
\begin{equation}\label{wigner}
d^{0}(\theta)=1,\qquad d^{1}_{mm'}(\theta)=\left(\begin{array}{ccc}
  \frac12(1+\cos\theta)&-\frac1{\sqrt2}\sin\theta&\frac12(1-\cos\theta)\\
  \frac1{\sqrt2}\sin\theta&\cos\theta&-\frac1{\sqrt2}\sin\theta\\
  \frac12(1-\cos\theta)&\frac1{\sqrt2}\sin\theta&\frac12(1+\cos\theta)
  \end{array}\right).
\end{equation}
The rows and columns in the spin-1 part of~(\ref{wigner}) are
labelled in the order $(+1,0,-1)$.

One obtains
\begin{eqnarray}\label{lepW}
(2q^{2})^{-1}L^{(q)}_{tt}&=&4\eps a_{\ell}^{2},\quad
L^{(q)}_{t\pm}\ =\ L^{(q)}_{\pm t}\ =\ 0,\quad
L^{(q)}_{t0}\ =\ L^{(q)}_{0t}\ =\ 0,\nn
(2q^{2})^{-1}L^{(q)}_{\pm \pm}
  &=&\tfrac12(1+\cos^{2}\theta)(v_{\ell}^{2}+a_{\ell}^{2}v_{q}^{2})
  \pm 2v_{\ell}a_{\ell}v \cos\theta +2\eps v_{\ell}^{2}\sin^{2}\theta,\nn
(2q^{2})^{-1}L^{(q)}_{00}&=&(v_{\ell}^{2}+a_{\ell}^{2}v_{q}^{2})
  \sin^{2}\theta +4\eps v_{\ell}^{2}\cos^{2}\theta,\nn
(2q^{2})^{-1}L^{(q)}_{\pm 0}&=&\tfrac{1}{2\sqrt{2}}
  \Big(\pm(v_{\ell}^{2}+a_{\ell}^{2}v_{q}^{2})\sin2\theta 
  -4v_{\ell}a_{\ell}v\sin\theta \mp 4\eps v_{\ell}^{2}\sin2\theta\Big)
  \,e^{\pm \,i\chi},\nn
L^{(q)}_{0\pm }&=& L^{(q)\dagger}_{\pm 0},\nn
(2q^{2})^{-1}L^{(q)}_{\pm \mp}
  &=&\Big(\tfrac12 (v_{\ell}^{2}+a_{\ell}^{2}v_{q}^{2})
  \sin^{2}\theta -2\eps v_{\ell}^{2}\sin^{2}\theta\Big)\,e^{\pm \,2i\chi}.
\end{eqnarray}
The corresponding expressions for the on-shell side lepton tensor
$L_{p^{2},\lambda_{Z}\,\lambda'_{Z}}(\cos\theta_{p})$ can again be obtained by
rotation. Note that in this case one has $\chi=0$ as is evident from
Fig.~\ref{planeszz}. 

The spin-1 and spin-0 projections of the neutral lepton tensor needed in the
main text are given by
\begin{eqnarray}
L_{1}(q^{2})&=&P_{1}^{\mu\nu}(q^{2}) L_{\mu\nu}(q^{2})\ =\ L_{U+L}(q^{2})
  \ =\ 4q^{2}\Big(v_{\ell}^{2}(1+2\eps)+a_{\ell}^{2}(1-4\eps)\Big),\nn
L_{0}(q^{2})&=&P_{0}^{\mu\nu}(q^{2}) L_{\mu\nu}(q^{2})\ =\ L_{tt}(q^{2})
  \ =\ 4q^{2}a_{\ell}^{2}2\eps,
\end{eqnarray}
where the spin-1 and spin-0 projectors read
\begin{equation}
P_{1}^{\mu\nu}(q^{2})=-g^{\mu\nu}+\frac{q^{\mu}q^{\nu}}{q^{2}},\qquad
P_{0}^{\mu\nu}(q^{2})=\frac{q^{\mu}q^{\nu}}{q^{2}}.
\end{equation}
Similar relations hold for the $p$ side.

\section{Helicity representation of the\\ charged-current lepton tensor}
\setcounter{equation}{0}\def\theequation{C\arabic{equation}}
The lepton tensors are given by
\begin{eqnarray}
\mbox {on-shell side}\,&:&
L^{(p)}_{\mu\nu}=\Tr\left(\slell^{-}\gamma_{\mu}(1-\gamma_{5})
  \bar\slnu\gamma_{\nu}(1-\gamma_{5})\right)\nn&&\qquad
  =8\left(\ell^{-}_{\mu}\bar\nu_\nu+\ell^{-}_{\nu}\bar\nu_{\mu}
  -\tfrac12 p^{2}g_{\mu\nu}-i\epsilon_{\mu\nu\rho\sigma}p^\rho\ell^{-\sigma}
  \right),\nn
\mbox {off-shell side}\,&:&
L^{(q)}_{\mu\nu}=\Tr\left(\slell^{+}+m_{\ell})\gamma_{\mu}
  (1-\gamma_{5})\slnu\gamma_{\nu}(1-\gamma_{5})\right)\nn&&\qquad
  =8\left(\ell^{+}_{\mu}\nu_{\nu}+\ell^{+}_{\nu}\nu_{\mu}
  -\tfrac12(q^{2}-m_{\ell}^{2})g_{\mu\nu}
  +i\epsilon_{\mu\nu\rho\sigma}q^\rho\ell^{+\sigma}\right).
\end{eqnarray}
The kinematics for the off-shell $q$ side is given by
\begin{eqnarray}
q^{\alpha}=\sqrt{q^{2}}(1;0,0,0)&&
\ell^{+\alpha}=\tfrac12\sqrt{q^{2}}(1+\eps;0,0,1-\eps)\qquad
\nu^{\alpha}=\tfrac12\sqrt{q^{2}}(1-\eps)(1;0,0,-1)\nn
\eps^{\mu}(0)=(0;0,0,1)&&
\eps^{\mu}(\pm)=\frac{1}{\sqrt{2}}(0;\mp1,-i,0)\qquad
\eps^{\mu}(t)=(1;0,0,0)
\end{eqnarray}
(for the on-shell $p$ side set $\eps=0$ and $q\to p$). The nonvanishing
components of the helicity representations of the lepton tensors can then be
evaluated to be
\begin{eqnarray}\label{hfnf}
\mbox {on-shell side}\,&:& \qquad \widehat L^{(p)}_{--}=8m^{2}_{W}\nn
\mbox {off-shell side}\,&:&
\widehat L^{(q)}_{++}=8q^{2}v_{\tau}\qquad \widehat L^{(q)}_{00}
=\widehat L^{(q)}_{0t}=\widehat L^{(q)}_{t0}
=\widehat L^{(q)}_{tt}=4m_{\tau}^{2}v_{\tau} 
\end{eqnarray}
where $v_{\tau}=1-m_{\tau}^{2}/q^{2}=1-\eps$. Note that the ratio of
helicity-flip and helicity-nonflip contributions are now given by e.g.\
$\widehat L^{(q)}_{00}/\widehat L^{(q)}_{++}=m_{\tau}^{2}/2q^{2}=\eps/2$.

As in the neutral-current case the components of the lepton tensor 
$L^{(q)}_{\lambda_{W}\,\lambda'_{W}}$ in the Higgs decay system ($z$ axis
along the $W^{+\ast}$ direction) needed in the evaluation of
Eq.~(\ref{angdis2}) can be obtained by rotating the components of the lepton
tensor $\widehat L^{(q)}_{\hat\lambda_{W}\,\hat\lambda'_{W}}$ in the helicity 
system according to (double indices are summed)
\begin{equation}
L^{(q)}_{\lambda_{W}\,\lambda'_{W}}(\cos\theta_{q},\chi)=\sum_{J,J'=0,1}
d^{J}_{\lambda_{W}\,\hat\lambda_{W}}(\theta_{q})\,
d^{J'}_{\lambda'_{W}\,\hat\lambda'_{W}}(\theta_{q})
e^{i(\lambda_{W}-\lambda'_{W})\chi}
\,\widehat L^{(q)}_{\hat\lambda_{W}\,\hat\lambda'_{W}}.
\end{equation}
One obtains (the rows and columns of the matrix are ordered in the sequence 
$(t,+1,0,-1)$)
\begin{eqnarray}\label{lt1}
\lefteqn{(2q^2v)^{-1} L_{\lambda_{W}\lambda'_{W}}(\cos\theta,\chi)\ =}\nn 
&&\left( \begin{array}{cccc} 
0 & 0 & 0 & 0 \\
0 & (1\mp\cos\theta)^2 & \mp \frac{2}{\sqrt{2}} (1\mp\cos\theta) \sin\theta 
e^{i\chi} & \sin^2\theta e^{2i\chi} \\
0 & \mp \frac{2}{\sqrt{2}} (1\mp\cos\theta) \sin\theta 
e^{-i\chi} & 2\sin^2\theta 
& \mp \frac{2}{\sqrt{2}} (1\pm\cos\theta) \sin\theta 
e^{i\chi} \\
0 & \sin^2\theta e^{-2i\chi} 
& \mp \frac{2}{\sqrt{2}} (1\pm\cos\theta) \sin\theta 
e^{-i\chi} & (1\pm\cos\theta)^2 \\
\end{array} \right)\nn&&
+\eps \left( \begin{array}{cccc} 
2 & - \frac{2}{\sqrt{2}} \sin\theta e^{i\chi} & 2 \cos\theta &
 \frac{2}{\sqrt{2}} \sin\theta e^{i\chi} \\
- \frac{2}{\sqrt{2}} \sin\theta e^{-i\chi} & 
 \sin^2\theta & 
- \frac{1}{\sqrt{2}} \sin 2\theta e^{i\chi} & 
-  \sin^2\theta e^{2i\chi} \\
2\cos\theta & 
- \frac{1}{\sqrt{2}} \sin 2\theta e^{-i\chi} & 
2\cos^2\theta & 
\frac{1}{\sqrt{2}} \sin 2\theta e^{i\chi} \\
\frac{2}{\sqrt{2}} \sin\theta e^{-i\chi} & 
-  \sin^2\theta e^{-2i\chi} & 
\frac{1}{\sqrt{2}} \sin 2\theta e^{-i\chi} & 
\sin^2\theta \\
\end{array} \right)
\end{eqnarray}
The upper/lower signs refer to the decays $W^{-}\to\ell^{-}\bar \nu_{\ell}$ and
$W^{+}\to\ell^{+}\nu_{\ell}$. The corresponding expressions for the
on-shell-side lepton tensor
$L^{(p)}_{\lambda_{W}\,\lambda'_{W}}(\cos\theta_{p})$ can again be obtained by
rotation. However, in this case one has $\chi=0$ as is evident from
Fig.~\ref{planesww}. Further, one has to use the lower signs in the
matrices~(\ref{lt1}).

The spin-1 and spin-0 projections of the charged lepton tensor are given by
\begin{eqnarray}
L^{W}_{1}(q^{2})&=&P_{1}^{\mu\nu}(q^{2})L^{(q)}_{\mu\nu}
  \ =\ 8q^{2}\left(1-\varepsilon\right)\left(1+\tfrac 12\varepsilon\right),
  \nonumber\\
L^{W}_{0}(q^{2})&=&P_{0}^{\mu\nu}(q^{2})L^{(q)}_{\mu\nu}
  \ =\ 8q^{2}\left(1-\varepsilon\right)\tfrac 12\epsilon.
\end{eqnarray}

\end{appendix}



\begin{thebibliography}{99}


\bibitem{:2012gk}
  G.~Aad {\it et al.\/} [ATLAS Collaboration],
  Phys.\ Lett.\ {\bf B716} (2012) 1

\bibitem{:2012gu}
  S.~Chatrchyan {\it et al.\/} [CMS Collaboration],
  Phys.\ Lett.\ {\bf B716} (2012) 30

\bibitem{Chatrchyan:2012jja}
  S.~Chatrchyan {\it et al.}  [CMS Collaboration],
  Phys.\ Rev.\ Lett.\ {\bf 110} (2013) 081803

\bibitem{Chatrchyan:2013mxa}
  S.~Chatrchyan {\it et al.}  [CMS Collaboration],
  Phys.\ Rev.\ {\bf D89} (2014) 092007

\bibitem{Aad:2013xqa}
  G.~Aad {\it et al.} [ATLAS Collaboration],
  Phys.\ Lett.\ {\bf B726} (2013) 120

\bibitem{Aad:2015rwa}
  G.~Aad {\it et al.}  [ATLAS Collaboration],
  arXiv:1503.03643 [hep-ex]

\bibitem{Choi:2002jk}
  S.Y.~Choi, D.J.~Miller, M.M.~M\"uhlleitner and P.M.~Zerwas,\\
  Phys.\ Lett.\ {\bf B553} (2003) 61

\bibitem{Kovalchuk:2008zz}
  V.A.~Kovalchuk,
  J.\ Exp.\ Theor.\ Phys.\ {\bf 107} (2008) 774

\bibitem{Gao:2010qx}
  Y.~Gao, A.V.~Gritsan, Z.~Guo, K.~Melnikov, M.~Schulze and N.V.~Tran,\\
  Phys.\ Rev.\ {\bf D81} (2010) 075022

\bibitem{DeRujula:2010ys}
  A.~De Rujula, J.~Lykken, M.~Pierini, C.~Rogan and M.~Spiropulu,\\
  Phys.\ Rev.\ {\bf D82} (2010) 013003

\bibitem{Bolognesi:2012mm}
  S.~Bolognesi, Y.~Gao, A.V.~Gritsan, K.~Melnikov, M.~Schulze, N.V.~Tran
  and A.~Whitbeck,
  Phys.\ Rev.\ {\bf D86} (2012) 095031

\bibitem{Avery:2012um}
  P.~Avery {\it et al.},
  Phys.\ Rev.\ {\bf D87} (2013) 055006

\bibitem{Sun:2013yra}
  Y.~Sun, X.F.~Wang and D.N.~Gao,
  Int.\ J.\ Mod.\ Phys.\ {\bf A29} (2014) 1450086

\bibitem{Buchalla:2013mpa}
  G.~Buchalla, O.~Cata and G.~D'Ambrosio,
  Eur.\ Phys.\ J.\ {\bf C74} (2014) 2798

\bibitem{Beneke:2014sba}
  M.~Beneke, D.~Boito and Y.M.~Wang,
  JHEP {\bf 1411} (2014) 028

\bibitem{Gainer:2014hha}
  J.S.~Gainer, J.~Lykken, K.T.~Matchev, S.~Mrenna and M.~Park,\\
  Phys.\ Rev.\ {\bf D91} (2015) 035011

\bibitem{Modak:2013sb}
  A.~Menon, T.~Modak, D.~Sahoo, R.~Sinha and H.Y.~Cheng,\\
  Phys.\ Rev.\ {\bf D89} (2014) 095021

\bibitem{Bhattacherjee:2015xra}
  B.~Bhattacherjee, T.~Modak, S.K.~Patra and R.~Sinha,
  arXiv:1503.08924 [hep-ph]

\bibitem{Zagoskin:2015sca}
  T.V.~Zagoskin and A.Y.~Korchin,
  arXiv:1504.07187 [hep-ph]

\bibitem{Ellis:2015daa}
  J.~Ellis,
  arXiv:1504.03654 [hep-ph]

\bibitem{Ellis:2015tba}
  J.~Ellis, M.~K.~Gaillard and D.~V.~Nanopoulos,
  arXiv:1504.07217 [hep-ph]

\bibitem{Djouadi:2015haa}
  A.~Djouadi,
  arXiv:1505.01059 [hep-ph]

\bibitem{Kadeer:2005aq}
  A.~Kadeer, J.G.~K\"orner and U.~Moosbrugger,
  Eur.\ Phys.\ J.\ {\bf C59} (2009) 27

\bibitem{Korner:1989ve}
  J.G.~K\"orner and G.A.~Schuler,
  Phys.\ Lett.\ {\bf B231} (1989) 306

\bibitem{Korner:1989qb}
  J.G.~K\"orner and G.A.~Schuler,
  Z.\ Phys.\ {\bf C46} (1990) 93

\bibitem{Gutsche:2015mxa}
  T.~Gutsche, M.A.~Ivanov, J.G.~K\"orner, V.E.~Lyubovitskij, P.~Santorelli
  and N.~Habyl,\\
  Phys.\ Rev.\ {\bf D91} (2015) 074001

\bibitem{Kniehl:2012rz}
  B.A.~Kniehl and O.L.~Veretin,
  Phys.\ Rev.\ {\bf D86} (2012) 053007

\bibitem{Peskin:1995ev}
  M.E.~Peskin and D.V.~Schroeder,
  ``An Introduction to quantum field theory,''
  Reading, USA: Addison-Wesley (1995) 842 p

\bibitem{Korner:2014bca}
  J.G.~K\"orner,
  arXiv:1402.2787 [hep-ph]

\bibitem{Cappiello:2011qc}
  L.~Cappiello, O.~Cata, G.~D'Ambrosio and D.N.~Gao,\\
  Eur.\ Phys.\ J.\ {\bf C72} (2012) 1872
  [Eur.\ Phys.\ J.\ {\bf C72} (2012) 2208]

\bibitem{Gevorkyan:2014waa}
  S.R.~Gevorkyan and M.H.~Misheva,
  Eur.\ Phys.\ J.\ {\bf C74} (2014) 2860

\bibitem{Korner:1987kd}
  J.G.~K\"orner and G.A.~Schuler,
  Z.\ Phys.\ {\bf C38} (1988) 511
  [Z.\ Phys.\ {\bf C41} (1989) 690]

\bibitem{Fischer:2002hn}
  M.~Fischer, S.~Groote, J.G.~K\"orner and M.C.~Mauser,
  Phys.\ Rev.\ {\bf D67} (2003) 113008

\bibitem{Fael:2013pja}
  M.~Fael, L.~Mercolli and M.~Passera,
  Phys.\ Rev.\ {\bf D88} (2013) 093011

\bibitem{Groote:2015}
  S.~Groote, J.G.~K\"orner and L.~Kaldam\"ae,
 ``Identical particle and lepton mass effects in the decay
  $H\to\tau^{+}\tau^{-}\tau^{+}\tau^{-}$'', in preparation

\bibitem{Groote:2012xr}
  S.~Groote, J.G.~K\"orner and P.~Tuvike,
  Eur.\ Phys.\ J.\ {\bf C72} (2012) 2177

\bibitem{Groote:2013xt}
  S.~Groote, J.G.~K\"orner and P.~Tuvike,
  Eur.\ Phys.\ J.\ {\bf C73} (2013) 2454

\bibitem{Aad:2015zhl}
  G.~Aad {\it et al.} [ATLAS and CMS Collaborations],
  Phys.\ Rev.\ Lett.\ {\bf 114} (2015) 191803

\bibitem{Agashe:2014kda}
  K.A.~Olive {\it et al.} [Particle Data Group Collaboration],
  Chin.\ Phys.\ {\bf C38} (2014) 090001

\bibitem{Gonzalez-Alonso:2014rla}
  M.~Gonzalez-Alonso and G.~Isidori,
  Phys.\ Lett.\ {\bf B733} (2014) 359

\bibitem{Keung:1984hn}
  W.Y.~Keung and W.J.~Marciano,
  Phys.\ Rev.\ {\bf D30} (1984) 248

\bibitem{Djouadi:2005gi}
  A.~Djouadi,
  Phys.\ Rept.\ {\bf 457} (2008) 1

\bibitem{Denner:2011mq}
  A.~Denner, S.~Heinemeyer, I.~Puljak, D.~Rebuzzi and M.~Spira,\\
  Eur.\ Phys.\ J.\ {\bf C71} (2011) 1753

\bibitem{Grau:1990uu}
  A.~Grau, G.~Panchieri and R.J.N.~Phillips,
  Phys.\ Lett.\ {\bf B251} (1990) 293

\bibitem{Zinonos:2014rma}
  Z.~Zinonos,
  arXiv:1409.0343 [hep-ex]

\bibitem{Sakurai:2014kha}
  Y.~Sakurai,
  arXiv:1409.2699 [hep-ex]

\bibitem{Jeans:2015vaa}
  D.~Jeans,
  arXiv:1507.01700 [hep-ex]

\bibitem{Aad:2015kxa}
  G.~Aad {\it et al.} [ATLAS Collaboration],
  arXiv:1506.05623 [hep-ex]

\bibitem{Bredenstein:2006rh}
  A.~Bredenstein, A.~Denner, S.~Dittmaier and M.M.~Weber,\\
  Phys.\ Rev.\ {\bf D74} (2006) 013004

\bibitem{Bredenstein:2006ha}
  A.~Bredenstein, A.~Denner, S.~Dittmaier and M.M.~Weber,
  JHEP {\bf 0702} (2007) 080

\end{thebibliography}
\end{document}